%% file: main.tex
\documentclass{article} %
\usepackage{colm2024_conference} %

\usepackage{microtype}
\usepackage[hidelinks]{hyperref} %
\usepackage{url}
\usepackage{booktabs}
\usepackage{xspace}
\usepackage{enumitem}
\usepackage[flushleft]{threeparttable}
\usepackage{makecell}
\usepackage{multirow}
\usepackage{xcolor, colortbl}
\usepackage{subcaption}
\usepackage{tikz}
\usepackage{wrapfig}
\usepackage{textcomp}  %
\usepackage{scalerel}  %
\usepackage{tipa}
\usepackage{pifont}
\usepackage[subtle]{savetrees}
\usepackage[most]{tcolorbox}
\usepackage{listings}
\usepackage{caption}
\usepackage[normalem]{ulem}

\colmfinalcopy
\begin{document}

\frenchspacing %

\definecolor{linkcolor}{HTML}{cc0000}
\input{macros}

\title{\tech{} \cat{}:\\
Demystifying LLM-based Software Engineering Agents}

\newcommand{\seperate}{{\ \ \ \ \ \ \ \ \ \ \ \ \ \ \ \ \ \ \ \ \ \ \ \ \ \ \ \ \ }}
\newcommand{\sseparate}{{\ \ }}

\author{Chunqiu Steven Xia\textsuperscript{$*$} 
\sseparate
Yinlin Deng\thanks{Contributed equally with author ordering decided by \href{https://senseis.xmp.net/?Nigiri}{\textit{Nigiri}}.}
\sseparate
\sseparate
\
Soren Dunn
\seperate
Lingming Zhang
\\[\bigskipamount]
University of Illinois Urbana-Champaign{ }\uiuc{}
\\[\bigskipamount]
\texttt{\{chunqiu2, yinlind2, sorend2, lingming\}@illinois.edu}
}

\maketitle

\begin{abstract}
Recent advancements in large language models (\llm{s}) have significantly advanced the automation of software development tasks, including code synthesis, program repair, and test generation. 
More recently, researchers and industry practitioners have developed various autonomous \emph{\llm agents} to perform end-to-end software development tasks. 
These agents are equipped with the ability to use tools, run commands, observe feedback from the environment, and plan for future actions. 
However, the complexity of these agent-based approaches, together with the limited abilities of current \llm{s}, raises the following question: 
\emph{Do we really have to employ complex autonomous software agents?}
To attempt to answer this question, we build \tech{} -- an \emph{agentless} approach to automatically resolve software development {issues}.
Compared to the verbose and complex setup of agent-based approaches, \tech employs a simplistic three-phase process of localization, repair, and patch validation, without letting the \llm decide future actions or operate with complex tools. 
Our results on the popular \swebenchlite benchmark show that surprisingly the simplistic \tech is able to achieve both the highest performance (\totalsolveperc{}, \totalsolve{} correct fixes) and low cost (\averagedollarcost) compared with all existing open-source software agents!
In fact, \emph{\tech has already been adopted by OpenAI as the go-to approach to showcase the real-world coding performance of both GPT-4o and the new OpenAI o1 models}.
Furthermore, we manually classified the problems in \swebenchlite and found problems with exact ground truth patches or insufficient/misleading issue descriptions.
As such, we construct \swebenchlitefiltered{} by excluding such problematic {issues} to perform more rigorous evaluation and comparison. 
Our work highlights the currently overlooked potential of a simplistic, {cost-effective} technique in autonomous software development.
We hope \tech will help reset the baseline, starting point, and horizon for autonomous software agents, and inspire future work along this crucial direction. 
We have open-sourced \tech at: \textcolor{linkcolor}{\url{https://github.com/OpenAutoCoder/Agentless}}

\end{abstract}

\input{secs/intro}

\input{secs/related}

\input{secs/approach}

\input{secs/experimental}
\input{secs/eval}
\input{secs/study}
\input{secs/conclusion}

\bibliographystyle{ACM-Reference-Format}
\bibliography{reference}

\end{document}

%% file: macros.tex
\newcommand{\tech}{\textsc{Agentless}\xspace} 

\newcommand{\reproduction}{reproduction\xspace}
\newcommand{\plausiblecorrect}{plausible\xspace}
\newcommand{\apr}{APR\xspace}

\newcommand{\totalbaselines}{26\xspace}

\newcommand{\sanitizedtotalproblem}{249\xspace}
\newcommand{\notenoughinfo}{10.0\%\xspace}
\newcommand{\exactpatch}{4.3\%\xspace}
\newcommand{\stepsinnl}{9.7\%\xspace}
\newcommand{\misleadingpatch}{5.0\%\xspace}

\newcommand{\totalsolveperc}{32.00\%\xspace}
\newcommand{\totalsolve}{96\xspace}
\newcommand{\averagedollarcost}{\$0.70\xspace}

\newcommand{\evileval}{\textsc{EvoEval}\xspace}
\newcommand{\humaneval}{\textsc{HumanEval}\xspace}
\newcommand{\evalplus}{\textsc{EvalPlus}\xspace}
\newcommand{\apps}{{APPS}\xspace}
\newcommand{\mbpp}{{MBPP}\xspace}
\newcommand{\pie}{{\textsc{pie}}\xspace}
\newcommand{\swebench}{SWE-bench\xspace}
\newcommand{\swebenchlite}{SWE-bench Lite\xspace}
\newcommand{\swebenchverified}{SWE-bench Verified\xspace}

\newcommand{\swebenchlitefiltered}{SWE-bench Lite-$S$\xspace}

\newcommand{\repostructureformat}{repository structure format\xspace}

\newcommand{\autocoderover}{AutoCodeRover\xspace}
\newcommand{\aider}{Aider\xspace}
\newcommand{\sweagent}{SWE-agent\xspace}
\newcommand{\coder}{CodeR\xspace}
\newcommand{\ibmagent}{IBM Research Agent-101\xspace}
\newcommand{\opencsgagent}{OpenCSG StarShip\xspace}
\newcommand{\bytedanceagent}{Bytedance MarsCode\xspace}
\newcommand{\amazonqagent}{Amazon Q Developer\xspace}
\newcommand{\opendevin}{OpenDevin\xspace}
\newcommand{\repounderstander}{RepoUnderstander\xspace}
\newcommand{\lingma}{Alibaba Lingma Agent\xspace}
\newcommand{\factorydroid}{Factory Code Droid\xspace}
\newcommand{\opendevincodeact}{OpenDevin+CodeAct v1.8\xspace}
\newcommand{\codestoryaide}{CodeStory Aide\xspace}
\newcommand{\mentatbot}{MentatBot\xspace}
\newcommand{\honeycomb}{Honeycomb\xspace}
\newcommand{\gru}{Gru\xspace}
\newcommand{\isoform}{Isoform\xspace}
\newcommand{\supercoder}{SuperCoder2.0\xspace}
\newcommand{\repograph}{RepoGraph\xspace}  %
\newcommand{\moatless}{Moatless\xspace}
\newcommand{\rag}{RAG\xspace}
\newcommand{\specrover}{SpecRover\xspace}
\newcommand{\masai}{MASAI\xspace}
\newcommand{\sima}{SIMA\xspace}
\newcommand{\appmapnavie}{AppMap Navie\xspace}
\newcommand{\toolstool}{Tools\xspace}
\newcommand{\solversolver}{Solver\xspace}
\newcommand{\composio}{Composio SWEkit\xspace}

\newcommand{\oone}{o1\xspace}
\newcommand{\codegen}{{CodeGen}\xspace}
\newcommand{\starcoder}{{StarCoder}\xspace}
\newcommand{\codegentwo}{{CodeGen2}\xspace}
\newcommand{\codex}{\textsc{Codex}\xspace}
\newcommand{\gptturbo}{GPT-3.5\xspace}
\newcommand{\gptfour}{GPT-4\xspace}
\newcommand{\gptfouro}{GPT-4o\xspace}
\newcommand{\gemini}{Gemini\xspace}
\newcommand{\claude}{Claude\xspace}
\newcommand{\claudetwo}{Claude 2\xspace}
\newcommand{\claudesonnet}{Claude 3.5 S\xspace}
\newcommand{\claudeopus}{Claude 3 Opus\xspace}
\newcommand{\claudehaiku}{Claude 3.5 Haiku\xspace}
\newcommand{\chatgpt}{GPT-3.5\xspace}
\newcommand{\phind}{{Phind-CodeLlama}\xspace}
\newcommand{\codellama}{CodeLlama\xspace}
\newcommand{\mistral}{Mistral\xspace}
\newcommand{\codetfp}{\textsc{CodeT5+}\xspace}
\newcommand{\wizardcoder}{{WizardCoder}\xspace}
\newcommand{\deepseekinstruct}{DeepSeek-Coder-Inst\xspace}
\newcommand{\deepseek}{DeepSeek-Coder\xspace}
\newcommand{\deepseekvonefive}{DeepSeek-Coder-1.5\xspace}
\newcommand{\codellamainstruct}{CodeLlama-Inst\xspace}
\newcommand{\phitwo}{Phi-2\xspace}
\newcommand{\openchat}{OpenChat\xspace}
\newcommand{\qwen}{Qwen-1.5\xspace}
\newcommand{\qwenb}{Qwen\xspace}
\newcommand{\palm}{PaLM-2\xspace}
\newcommand{\phindllamatwo}{Phind-CodeLlama-2\xspace}
\newcommand{\mistralinstruct}{Mistral-Inst\xspace}
\newcommand{\mixtralinstruct}{Mixtral-Inst\xspace}
\newcommand{\codemillenials}{Code Millenials\xspace}
\newcommand{\xwincoder}{XwinCoder\xspace}
\newcommand{\stablecode}{stable-code\xspace}
\newcommand{\gemma}{Gemma\xspace}
\newcommand{\speechlesscodellama}{Speechless-CL\xspace}
\newcommand{\starcodertwo}{StarCoder2\xspace}
\newcommand{\magicoder}{Magicoder\xspace}
\newcommand{\lingmaswegpt}{Lingma SWE-GPT 72b\xspace}

\newcommand{\groundtruth}{ground-truth\xspace}
\newcommand{\gt}{ground-truth\xspace}
\newcommand{\Gt}{Ground-truth\xspace}
\newcommand{\llmfull}{large language model\xspace}
\newcommand{\llm}{LLM\xspace}
\newcommand{\nlpfull}{Natural Language Processing\xspace}
\newcommand{\nlp}{NLP\xspace}

\newcommand{\lingming}[1]{{\color{red}\bfseries [Lingming: #1]}}
\newcommand{\steven}[1]{{\color{blue}\bfseries [Steven: #1]}}  
\newcommand{\yinlin}[1]{{\color{olive}\bfseries [Yinlin: #1]}}  
\newcommand{\soren}[1]{{\color{violet}\bfseries [Soren: #1]}}  

\newcommand{\passat}[1]{\textls[-25]{pass{@}\(#1\)}\xspace}
\newcommand{\parabf}[1]{\vspace{.03in}\noindent \textbf{#1}}
\newcommand{\CodeIn}[1]{{\small \texttt{#1}}}
\newcommand{\Comment}[1]{}
\newcommand{\edit}[2]{{\color{red}\sout{#1}}{\color{blue}#2}}

\newcommand{\eg}{e.g.,\xspace}
\newcommand{\ie}{i.e.,\xspace}

\definecolor{yucky}{HTML}{a64d79}
\newcommand*\circled[1]{\scalebox{0.8}{\tikz[baseline=(char.base)]{
\node[anchor=text, shape=circle,fill=yucky, inner sep=0pt, minimum size=1.2em] (char) {\footnotesize \textcolor{white}{#1}};}}}

\newcommand{\eating}{\scalerel*{\includegraphics{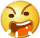}}{\textrm{C}}\xspace}
\newcommand{\smallmonkey}{\scalerel*{\includegraphics{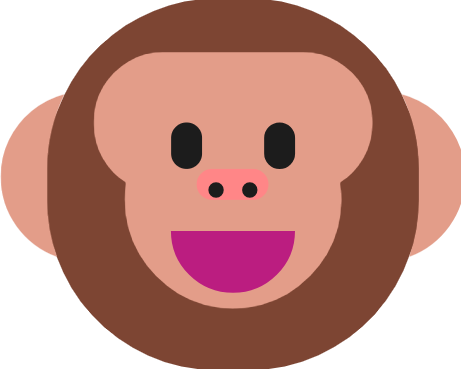}}{\textrm{C}}\xspace}
\newcommand{\cat}{\scalerel*{\includegraphics{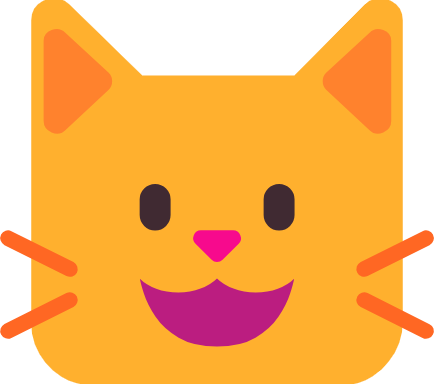}}{\textrm{C}}\xspace}
\newcommand{\grinningcat}{\scalerel*{\includegraphics{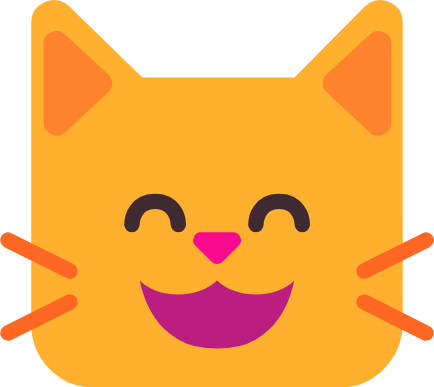}}{\textrm{C}}\xspace}
\newcommand{\wearycat}{\scalerel*{\includegraphics{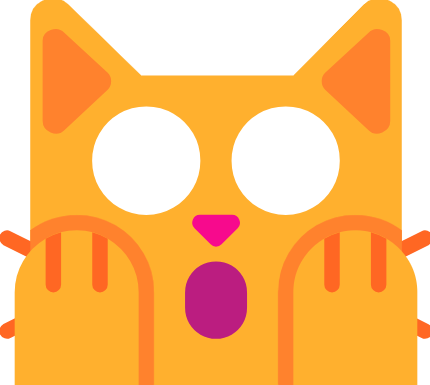}}{\textrm{C}}\xspace}
\newcommand{\smugcat}{\scalerel*{\includegraphics{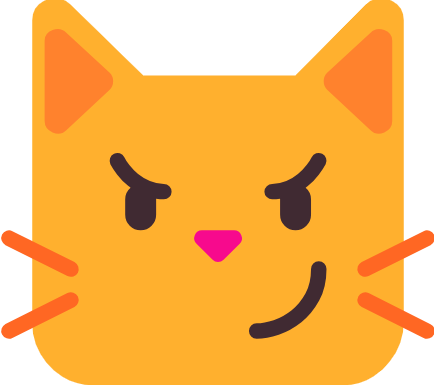}}{\textrm{C}}\xspace}
\newcommand{\questionmark}{\scalerel*{\includegraphics{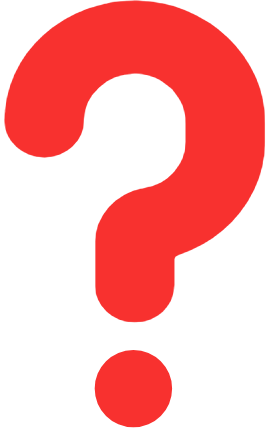}}{\textrm{C}}\xspace}

\newcommand{\lock}{\scalerel*{\includegraphics{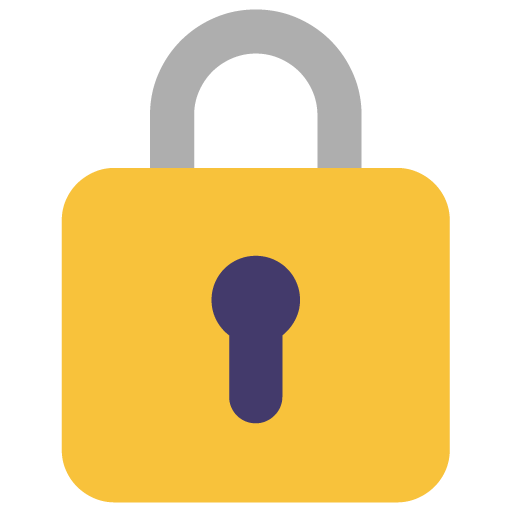}}{\textrm{C}}\xspace}

\newcommand{\uiuc}{\scalebox{1}{\scalerel*{\includegraphics{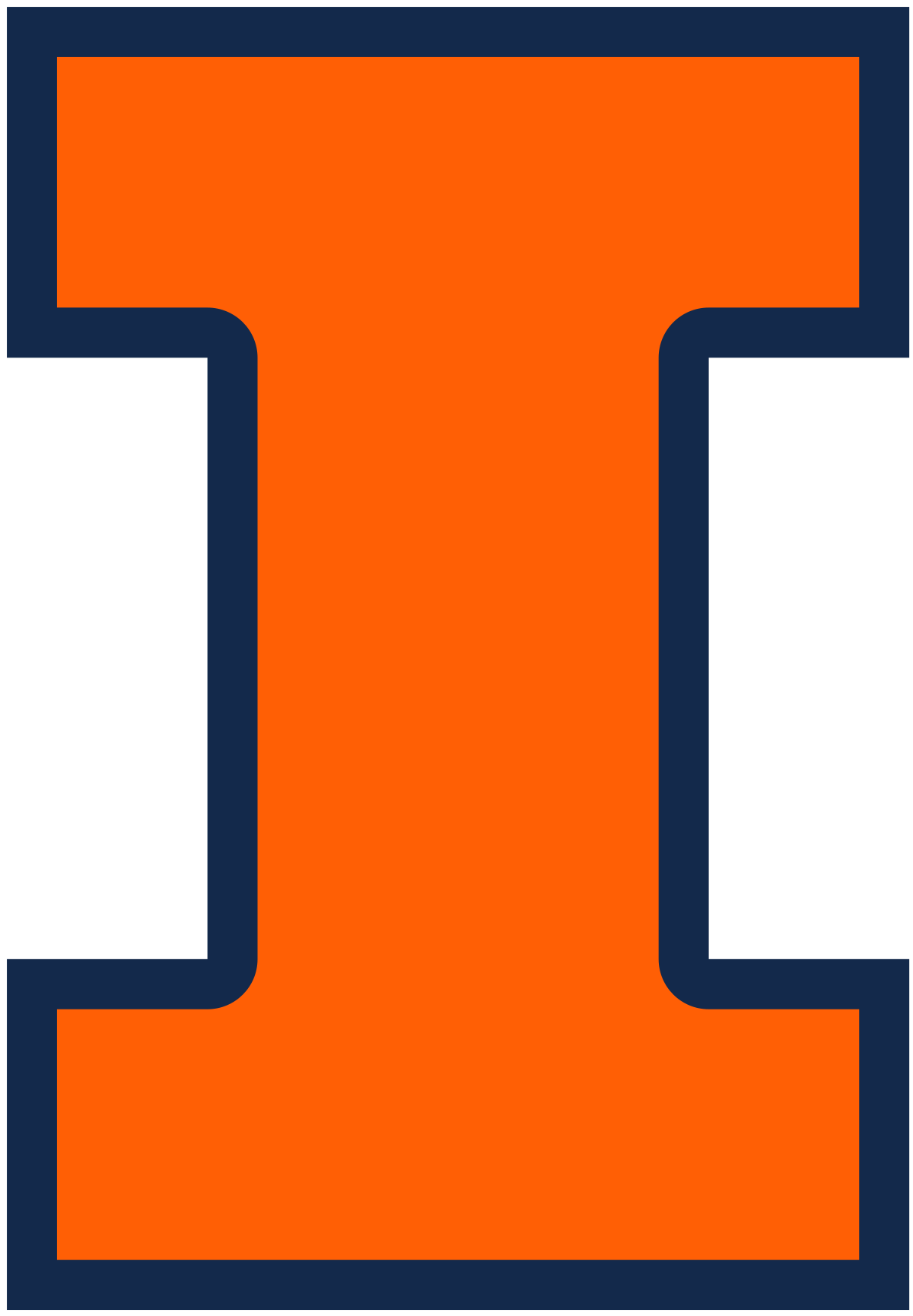}}{\textrm{C}}}\xspace}
\newcommand{\openai}{\scalerel*{\includegraphics{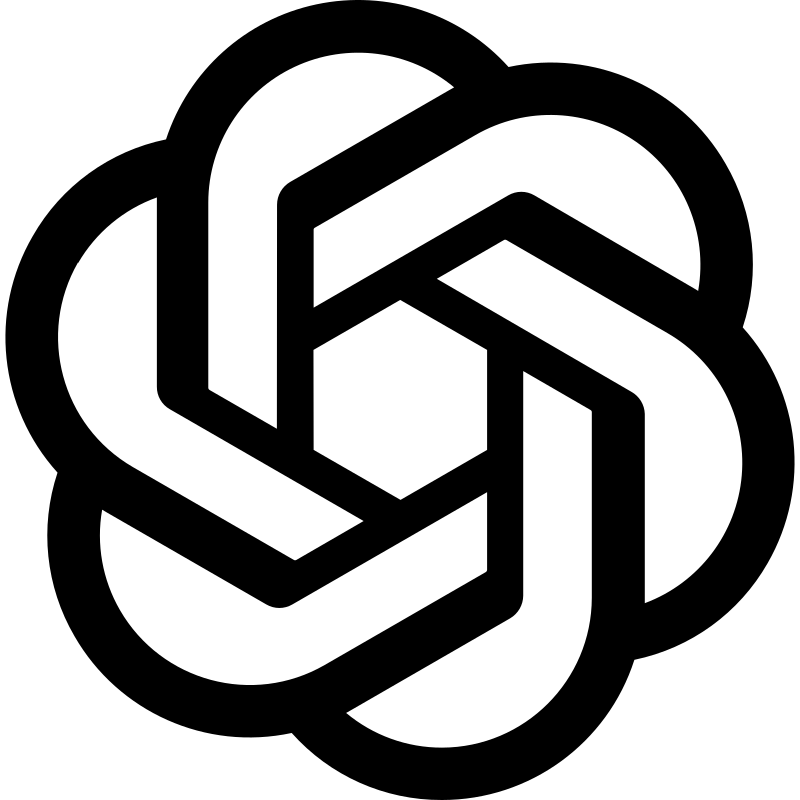}}{\textrm{C}}\xspace}
\newcommand{\anthropic}{\scalebox{1}{\scalerel*{\includegraphics{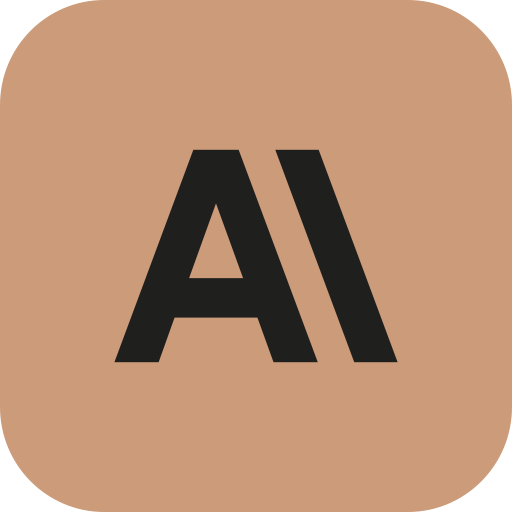}}{\textrm{C}}}\xspace}

\newcommand{\cmark}{\ding{51}}
\newcommand{\xmark}{\ding{55}}

\lstset{
  basicstyle=\ttfamily\small,
  columns=fullflexible,
  breaklines=true,
  postbreak=\mbox{$\hookrightarrow$\space},
  rulecolor=\color{black},
}
\definecolor{codegreen}{rgb}{0,0.6,0}
\lstdefinestyle{mystyle}{  
    commentstyle=\color{codegreen},
    keywordstyle=\color{blue},
    basicstyle=\ttfamily\small,
    breakatwhitespace=false,        
    breaklines=true,                 
    captionpos=b,                    
    keepspaces=true,                 
    showspaces=false,                
    showstringspaces=false,
    showtabs=false,                  
    tabsize=2
}
\lstset{style=mystyle}

%% file: secs/intro.tex
\section{Introduction}

Large language model{s} (\llm{s}) have become the go-to default choice for code generation~\citep{codex,austin2021program,starcoder,magicoder}. 
State-of-the-art \llm{s} like \gptfour~\citep{gptfour} and \claudesonnet{}onnet~\cite{claudethreefive} have demonstrated their prowess in being able to synthesize code snippets based on given user description. 
However, compared to the main evaluation setting of simple, self-contained problems, applying \llm{s} on repository-level software engineering tasks has been understudied. 
Software engineering tasks like feature addition, program repair, and test generation require an in-depth understanding of not only information within files, containing thousands of lines of code, but also repository-level dependencies across files. 

Recently, to address the gap and evaluate the ability of tools to automatically solve real-world software engineering problems, the popular \swebench~\cite{swebench} benchmark has been developed. 
In \swebench, each problem consists of a real-world GitHub issue description and the corresponding Python repository.
The task is to modify the repository to resolve the issue, either fixing a bug or introducing a new feature.
Recently, the authors have published a subset of the benchmark -- \swebenchlite~\cite{swebenchlite} (300 problems) that performs further filtering and focuses on bug fixing issues. 

To solve the challenging real-world software development problems from \swebench, {inspired by the Devin AI Software Engineer~\cite{devinwebpage}, there has been a significant body of work from both academia and industry focusing on developing \emph{agent-based} approaches~\cite{autocoderover, aidar, sweagent, coder, repounderstander, bouzenia2024repairagent}}. 
While there is not a fixed definition for agent-based approaches, they generally equip \llm{s} with a set of tools and allow agents to iteratively and autonomously perform actions, observe feedback, and plan future steps~\cite{liu2024largesurvey}. 
Example tools can include the ability to open/write/create files, search for code lines, run tests, and execute shell commands.  
In each attempt to solve a problem, agent-based approaches will have multiple turns, where each turn consists of performing an action.
Subsequent turns {depend} on previous actions and the feedback information the agent receives from the environment.

At first glance, agent-based approaches appear to be a natural and straightforward way to tackle software development tasks. 
After all, human developers also perform similar actions and use feedback to plan future steps. 
However, the disparity between human and current \llm abilities leads to the following limitations of agent-based approaches:

\begin{itemize}[noitemsep, leftmargin=*, topsep=0pt]
    \item \textbf{Complex tool usage/design.}
    To utilize tools, current agent-based approaches apply an abstraction layer between the agent and the environment. 
    Examples are mapping real actions to API calls so that agents can use tools by outputting an API call instruction.
    However, such abstractions and API call specifications require careful design of input/output formats and can easily lead to incorrect or imprecise tool design/usage,
    especially for more complex action spaces. 
    Given the{ iterative} nature of agent-based approaches, where current action/plan depends on previous turns, incorrectly or imprecisely defining/using a tool can both reduce performance and incur additional cost in wasted \llm queries. 
    \item {\textbf{Lack of control in decision planning.}}
    In addition to using tools, current agent-based approaches also delegate the decision-making process to the agents{}, allowing them to decide when and what action to perform. 
    The agents decide the current action to take based on previous actions taken and the feedback provided by the environment, often with minimal checks to ensure the action taken make sense.
    Due to the large possible action space and feedback response, it can be extremely easy for autonomous agents to become confused and perform sub-optimal explorations.
    Furthermore, to solve an issue, an agent can take upwards of 30 or 40 turns, which makes it extremely difficult to both understand the decisions made by the agents and also debug the exact turns where an incorrect decision is made. 
    \item \textbf{Limited ability to self-reflect.}
    Existing agents struggle with the capability to perform self-reflection~\cite{olausson2023self, huang2024large}.
    That is to say they tend to take all information/feedback and do not know how to filter out or correct irrelevant, incorrect, or misleading information~\cite{shi2023large, zhang2023siren}. 
    The limited ability to self-reflect means that an incorrect step can be easily amplified and negatively affect all future decisions made by the agent.
    
\end{itemize}

In this paper, we advocate that instead of rushing to develop increasingly complex \llm agent-based approaches and tools for software development (which can also be non-trivial to use or replicate due to the fully autonomous setup), we should first take a step back and ask the following introspective question: \emph{Do we really have to employ complex autonomous software agents?}

\parabf{Our work.} We set out to answer this important question by building \tech{} -- an \emph{agentless} approach to automatically resolve software development {issues}.
To solve each issue, \tech follows a simple three phase process: localization, repair, and patch validation. 
In the localization process, \tech employs a hierarchical process to first localize the fault to specific files, then to relevant classes or functions, and finally to fine-grained edit locations.
\tech's localization process make uses of both \llm-based localization as well classic information-retrieval-based localization idea~\cite{zhou2012should}.
To perform repair, \tech takes the localized edit locations and generates multiple candidate patches in a simple diff format. 
At the same time, \tech generates \reproduction tests that can reproduce the original error and help with candidate patch selection.
Finally, \tech re-ranks all remaining patches and selects one to submit in order to fix the issue. 
While \tech leverages \llm{s} to perform each detailed task, unlike prior complex agent-based tools, \tech{} does not allow \llm{s} to \emph{autonomously} decide future actions or operate with any complex tools.
Our deliberate choice {to avoid} using agents not only allows \tech to have a simplistic and straightforward design that {is easy to understand}, but also helps avoid the above-mentioned limitations of \llm agents in software development. 
We evaluate \tech on the popular \swebenchlite~\citep{swebenchlite} benchmark and demonstrate that \tech not only achieves the highest performance (\totalsolveperc{}) {among} all open-source approaches, but {also} does so at a fraction of the cost! 

Furthermore, we performed a fine-grained manual analysis on the \swebenchlite dataset and {classified} all its problems into different categories across dimensions like problem description, ground truth patch, and {bug} location information.
Surprisingly, we observed that \swebenchlite contains problems with exact ground truth patch in the description (\exactpatch), problems with missing critical information needed to solve the issue (\notenoughinfo), and problems that include misleading solutions in the issue description (\misleadingpatch).
Recognizing these issues, we built \swebenchlitefiltered{},
which removes such problematic problems, and serves as a more rigorous benchmark to evaluate the ability to solve real-world software development {challenges}.
Overall, {in an era focused on achieving} top placements on leaderboards, our work highlights the overlooked potential of a simplistic, cost-effective technique in autonomous software development.
We hope \tech will help reset the baseline, starting point, and horizon for autonomous software agents, and inspire future work along this crucial direction.

\parabf{Contributions.} In this work, we make the following contributions:

\begin{itemize}[noitemsep, leftmargin=*, topsep=0pt]
 \item \textbf{An \emph{agentless} approach.} We propose \tech, an \emph{agentless} approach to automatically solve software development problems.
 \tech leverages LLM-empowered prompting-based and embedding-based retrieval to perform hierarchical localization.
 During repair, \tech{} samples multiple candidate patches in a simple diff format for efficient patch generation.
 \tech{} {then} generates \reproduction tests to verify {that} the issue has been resolved. 
 Finally, \tech leverages both regression {tests} and generated \reproduction tests to select the final submission patch.
 \item \textbf{Extensive evaluation.} We evaluate \tech on the popular \swebenchlite dataset comparing against state-of-the-art agent-based approaches. 
 Our results demonstrates that \tech is able to achieve higher performance  (\totalsolveperc{}, \totalsolve correct fixes) than all open-source approaches, with comparably low cost as well.
 This shows the previously overlooked potential of a simplistic technique in autonomous software development. 
 Additionally, \emph{\tech has already been adopted by OpenAI as the go-to approach to showcase the real-world coding performance of \gptfouro~\cite{openaiverified} as well as the new \oone model family~\cite{openaioonesystemcard}}
 We further perform a rigorous ablation study to understand the effectiveness of different components of \tech on the final performance.
 \item \textbf{\swebenchlitefiltered{} benchmark.} We performed manual classifications on the problems in the popular \swebenchlite dataset.
 We found that there are problematic problems with unclear, misleading issue descriptions, as well as problems that contain exact ground truth patches. 
 To address these issues, we constructed a filtered dataset of \swebenchlitefiltered{} that excludes such problematic issues{, enabling} more rigorous evaluation and comparison. 
 This has also been confirmed recently by OpenAI, which acknowledged our benchmark and released \swebenchverified~\cite{openaiverified} along the same direction.
\end{itemize}

%% file: secs/related.tex
\section{Background and Related Work}

\subsection{Agent-based Software Engineering}

With the emergence and popularity of agent-based frameworks~\cite{xi2023rise}, recently researchers and industry practitioners have begun developing agent-based approaches to solve software engineering tasks. 
Devin~\cite{devinwebpage} (and OpenDevin~\cite{opendevin}, open-source alternative), is one of the first end-to-end \llm agent-based framework.
Devin uses agents to first perform planning based on user requirement, then allows the agent to use file editor, terminal, and web search engine tools to iteratively perform the task.
\sweagent~\cite{sweagent} designs a custom agent-computer interface (ACI) that allows the \llm agent to interact with the repository environment with actions such as reading, editing files, and running bash commands. 
\aider~\cite{aidar} first provides a detailed repository map constructed with static and call graph analysis to the \llm to localize the files that require editing; then it generates a simple diff format as the editing patch and uses regression testing to verify if the patch is plausible.
\moatless~\cite{moatless} is another open-source agent tool that obtains relevant code locations by providing the agent with both code search tools as well as retrieval methods using \llm-constructed queries. 
Similar to \aider, \moatless also generates a simple diff format as the final submitted patch. 
\autocoderover~\cite{autocoderover} further provides the \llm agent with specific code search APIs (\eg searching methods in a certain class) to iteratively retrieve code context and {locate the bug locations}. 
\specrover~\cite{specrover} later improves over \autocoderover and targets specifications (i.e., inferring the intended program behavior) by generating function summaries and also feedback messages during specific agent steps. 
Furthermore, \specrover also attempts to generate \reproduction tests to reproduce the original issue used to select the final patch. 
In addition to these highlighted examples, there has been a plethora of other agent-based approaches developed in both open-source~\cite{aidar, repograph, appmapnavie, zhang2024diversity} and close-source/commercial products~\cite{bouzenia2024repairagent, coder, repounderstander, lingma, factorydroid, ibmagent, opencsgstarship, marscode, amazonqdeveloper, supercoder, gru, isoform, mentatbot, honeycomb, codestoryaide}.

Compared to these agent-based techniques, \tech offers a simplistic, interpretable, and {cost-effective} solution to tackle real-world software engineering issues. 
Different from agent-based tools, \tech contains well-defined stages of localization, repair, and patch validation without letting the \llm agent decide future actions or use complex tools.
\tech demonstrates for the first time that an \emph{agentless} approach can achieve very competitive performance, without the additional baggage of having to provide excessive tools or model complex environment behavior/feedback.

\subsection{Fault Localization and Program Repair}

Fault localization (FL)~\cite{wong2016survey} techniques aim to identify the suspicious locations (e.g., statements or methods) in source code related to bugs.
Dynamic FL techniques mainly include spectrum-based FL (SBFL)~\cite{jones2005empirical, abreu2007accuracy, abreu2009practical} and mutation-based FL (MBFL)~\cite{papadakis2015metallaxis, moon2014ask, lou2020can}. SBFL typically computes source code locations primarily covered by failing tests as more suspicious than locations primarily covered by passing tests. MBFL further improves upon that to additionally consider the impact of each source code location on the test outcomes (measured using mutation testing~\cite{papadakis2019mutation}). Besides dynamic techniques, researchers have also proposed to directly leverage information retrieval (IR) techniques~\cite{singhal2001modern} for static FL. 
Such IR-based techniques~\cite{wang2015evaluating, saha2013improving, wang2014version} formulate FL as a search problem and compare the textual similarity between code elements and the bug report (i.e., query).
Moreover, learning-based techniques have also been proposed to leverage machine learning to combine multiple sources of dynamic/static information, including DeepFL~\cite{li2019deepfl}, FLUCCS~\cite{sohn2017fluccs}, and TRANSFER~\cite{meng2022improvingfl}. 
Recently, researchers have proposed \llm-based FL~\cite{yang2024llmao, wu2023large, qin2024agentfl, kang2024quantitative}, which leverages the powerful code and natural language understanding of modern \llm{s} to directly localize bugs. Meanwhile, most such \llm-based techniques either cannot perform repository-level FL due to the limited context window of \llm{s}~\cite{yang2024llmao}, or rely on complicated/costly agentic design to navigate through the codebase~\cite{kang2024quantitative, qin2024agentfl}. 
In contrast, \tech employs a simplistic hierarchical FL process (based on both \llm{s} and IR) to efficiently compute the fine-grained edit locations.

After localizing the bug, the next step is to perform repair. 
Automated program repair~\cite{gazzola2019aprsurvey} (APR) has been widely studied to automatically generate patches for bugs. 
Traditional APR techniques can be categorized as template-based~\cite{liu2019tbar, ghanbari2019prapr}, heuristic-based~\cite{legoues2012genprog, le2016hdrepair}, and constraint-based~\cite{mechtaev2016angelix, long2015spr} tools. 
While effective, traditional \apr tools suffer from scalability issues and are limited by their patch variety.
As such, researchers have proposed learning-based \apr tools either by training NMT (neural machine translation) models~\cite{jiang2021cure, li2020dlfix, chen2018sequencer, zhu2021recoder} or using pre-trained \llm{s} to perform repair~\cite{alpharepair, xia2023repairstudy, kolak2022patch, zhang2024systematicapr}.
Specifically, \llm-based APR tools, which sample multiple candidate patches per bug, have been shown to be the state-of-the-art due to the powerful coding capability of modern \llm{s}~\cite{xia2023repairstudy}. More recently, agent-based \apr techniques have also been proposed~\cite{chatrepair, bouzenia2024repairagent, hidvegi2024cigar, chen2024flakiness, zhang2024acfix}. 
Inspired by existing \llm-based \apr tools, \tech samples multiple candidate patches per issue to maximize the chance of generating a correct fix. Different from most \llm-based \apr techniques, \tech generates patches using a simple diff format~\cite{aidar} to avoid generating the complete code and instead focus on producing cost-efficient small edits, increasing the reliability and accuracy of patch generation (less chances for hallucination).
Furthermore, different from the simplistic bugs studied in most prior work~\cite{xia2023repairstudy}, \tech targets complex repository-level issues spanning multiple locations. 

\subsection{\llm-based Test Generation}

In addition to localizing and repairing bugs, another research area that has been adopting \llm is test generation. 
One area of test generation is fuzz testing~\cite{zeller2019fuzzing}, also known as fuzzing, to generate large amounts of inputs in order to expose bugs in systems. 
Researchers have applied \llm{s} to perform fuzzing in domains such as DL libraries~\cite{titanfuzz, deng2023fuzzgpt}, OS Kernel~\cite{yang2023kernelgpt, oliinyk2024fuzzing}, compilers~\cite{fuzz4all,yang2023whitebox, ou2024mutators}, network protocols~\cite{meng2024large}, and mobile applications~\cite{liu2024make}. 
\llm-based fuzzers have demonstrated their impact by detecting many bugs not found by traditional fuzzers as well as unlocking new fuzzing domains.
Besides fuzzing, researchers have also proposed to leverage \llm{s} for unit test generation to test individual software units (e.g., methods/classes)~\cite{liu2024largesurvey}, such as CodeMosa~\cite{lemieux2023codamosa}, ChatTester~\cite{yuan2024evaluating}, TestPilot~\cite{schafer2023testpilot}, and CoverUp~\cite{pizzorno2024coverup}.

Bug reproduction is a critical step in investigating bug reports~\cite{jin2012bugredux}, and is integrated into many recent software engineering agents~\cite{opendevin,specrover,honeycomb,mentatbot,factorydroid,sweagent,masai}.
For example, \specrover~\cite{specrover} begins by generating a test to reproduce the issue described in the bug report; the test then guides the context retrieval and patching process.%
Unlike such agent-based approaches, which generate a reproduction test and rely on an LLM agent to decide whether the test is correct, \tech { simply executes multiple sampled tests} and verifies if the execution results indicate the issue has been reproduced.

%% file: secs/approach.tex
\section{\tech \protect\cat Approach }

\begin{figure}
    \centering
    \includegraphics[width=1\columnwidth]{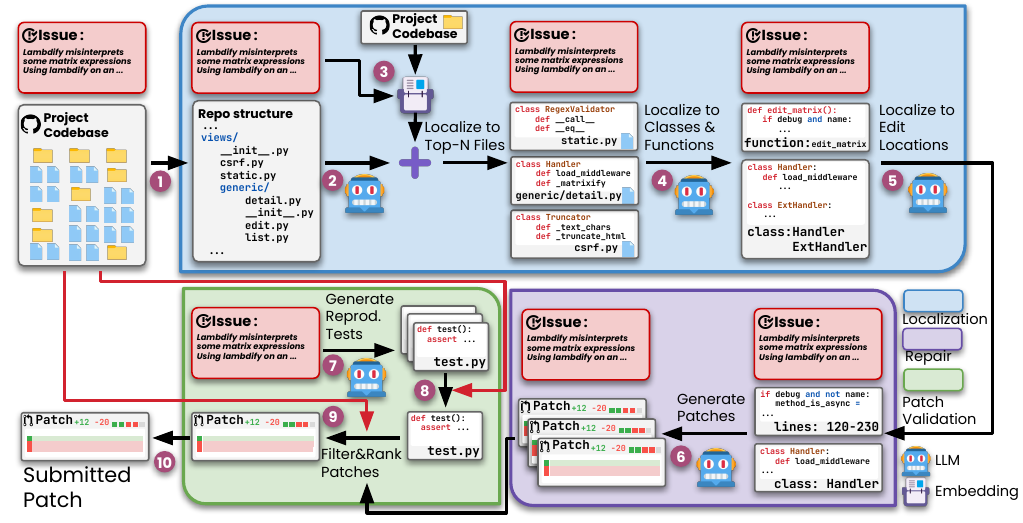}
    \caption{Overview of \tech.}
    \label{fig:overview}
\end{figure}

Figure~\ref{fig:overview} shows the overview of \tech, consisting of three phases: \textbf{localization}, \textbf{repair}, and \textbf{patch validation}. 
We first take in the issue description and the existing project codebase as input. 
Then, we begin our hierarchical localization process by turning the project codebase into a tree-like structure that {illustrates} the relative location of each file in the project \circled{1}.
Next, using this repository structure along with the original issue description, we {prompt} the \llm to localize and rank the top N most suspicious files that {likely require} editing to solve the issue \circled{2}. 
Since our repository structure format does not contain detailed source code information, we additionally retrieve files with most relevant code snippets with the issue description{ using embedding-based retrieval}  \circled{3}. 
We then combine the retrieved files with the \llm-localized files to obtain the final list of suspicious files. 
However, not all contents in each file need to be modified. 
As such, we provide a skeleton for each file (\ie a list of declaration headers of the classes and functions) and ask the \llm to output a specific list of classes and functions that we should examine more closely to fix the bug \circled{4}.
We then provide the complete code content of the previous locations and ask the \llm to finalize a smaller set of edit locations (\ie classes, functions, or even specific lines) \circled{5}.
For the repair phase, we provide the code snippets at these edit locations together with the issue description and {prompt} the \llm to sample multiple patches to solve the issue \circled{6}.
Next, we enter the patch validation phase, where we first ask the \llm to sample multiple \reproduction tests that aim to replicate the original issue \circled{7}, and then select the optimal one based on actual execution results on the original codebase \circled{8}. 
\tech uses the \reproduction test along with existing regression tests for patch ranking/selection \circled{9}.
Finally, \tech selects the top-ranked patch as the final patch for submission \circled{10}.
We now describe the steps in each of \tech{}'s phases in more detail.

\subsection{Localization \protect\wearycat}

To fix or implement a new feature, the first step is to obtain the locations in the source code, as without the correct locations, it can be impossible to provide the right edits. 
The difficulty lies in the fact that there could be hundreds of files
with thousands of lines of code each in a repository, whereas the correct locations to edit are only a few selected lines or functions.
\tech addresses this by using a simple three-step hierarchical localization process: 1) localize to {suspicious} files; 2) localize each selected files into relevant classes{, functions, and variables;} 3) localize to code edit locations.

\subsubsection{Localize to suspicious files.} 
\label{sec:localize_suspicious_files}
First, \tech narrows down potential locations to specific suspicious files. 
Instead of providing the complete code snippet for each file, \tech constructs {a concise representation of} the repository's file and directory structure, similar to the Linux \CodeIn{tree} command. 
We refer to this as the \textbf{\repostructureformat},
which begins with the root folder of the repository and organizes code files or folder names. 
Files and folders at the same directory level are aligned vertically, and files/folders in sub-directories are indented. 
We recursively traverse the entire repository to obtain the structure, which will be used as input for the \llm. 
The \repostructureformat provides the necessary file path{s alongside} the neighboring file names {}{to maintain organizational information in the original codebase}.
\tech then inputs the processed repository structure along with the original issue description to an \llm, and requests it to identify {a} list of the top N suspicious files{ that need {}{further inspection or} modification} to {resolve} the issue. 

To compliment the prompting-based localization (using file names only), \tech also uses a simple embedding-based retrieval method to {identify} additional suspicious files. 
However, instead of embedding all files in the repository, \tech first {filters out} irrelevant folders. 
This is done by providing the previously described repository structure and asking the \llm to produce a list of \emph{irrelevant} folders that do not need to be further inspected or modified to resolve the issue.
After removing all files from these irrelevant folders, \tech{} divides each remaining file into chunks of code segments and computes the embedding for each chunk using an embedding model. 
\tech then embeds the original issue description (i.e., the query) and computes the cosine similarity between the resulting query embedding and each chunk embedding to retrieve a list of relevant files that contain code segments with the highest similarity to the query. 
Finally, \tech combines the files obtained via prompting with those retrieved via embedding by selecting top N most common files localized by both, resulting in a final list of relevant files.

\begin{wrapfigure}{r}{0.35\textwidth}
  \begin{center}
    \includegraphics[width=0.35\textwidth]{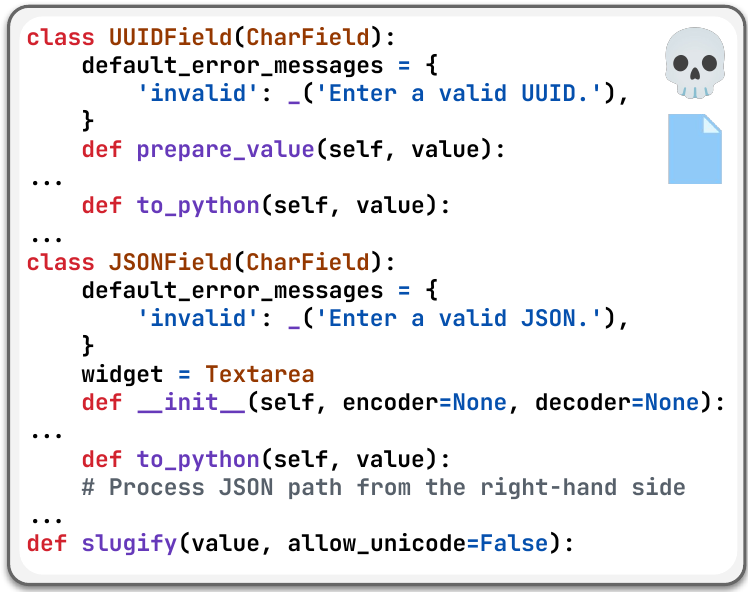}
  \end{center}
  \caption{File skeleton format.}
  \label{fig:skeleton_example}
\end{wrapfigure}

\subsubsection{Localize to related elements.}
\label{sec:localize_related_elements}
After obtaining the list of suspicious files, \tech{} {proceeds} to the second part of the localization process: localize the related{ elements} within these files.
Directly providing the complete context of all files can be large.
As such, \tech builds a compressed format of each file that contains the list of class, function, or variable declarations.
We refer to this format as the \textbf{skeleton format}, with an example shown in Figure~\ref{fig:skeleton_example}. 
In the skeleton format, we provide only the {headers} of the classes and functions in the file.
For classes, we further include any class fields and methods (signatures only).
Additionally, we also keep comments in the class and module level to provide further information. 
Compared to providing the entire file context to the model, the skeleton format is a much more concise representation, especially when the file contains thousands of lines, making it impractical/costly to process all at once with existing \llm{s}.
We provide the skeleton of all suspicious files to the \llm at one time in a single prompt, enabling the model to comprehensively analyze the pertinent information and decide the most relevant elements.
Using this input, we prompt the \llm to provide a list of related classes and functions that one should examine to fix the provided issue.

\begin{wrapfigure}{r}{0.45\textwidth}
  \begin{center}
    \includegraphics[width=0.45\textwidth]{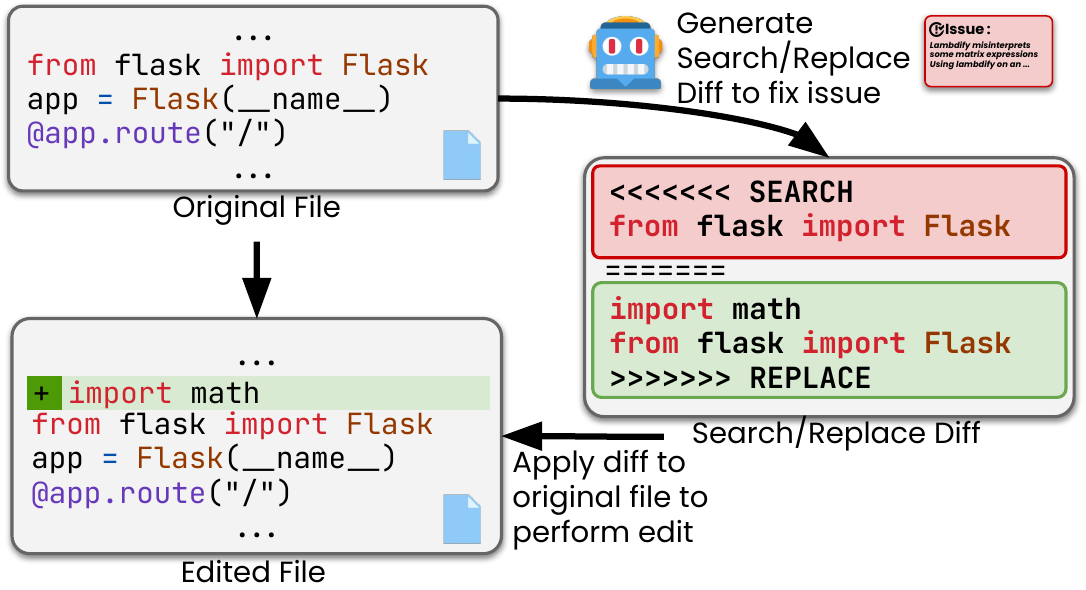}
  \end{center}
  \caption{Search/Replace edit format.}
  \label{fig:diff_format}
\end{wrapfigure}

\subsubsection{Localize to edit locations.}
The previous localization step provided us with a list of related code elements; since we localize top N suspicious files, these localized related code elements could be from different files.
We now directly provide the code content from these elements to the \llm and ask it to localize specific edit locations.
Compared to using the entire file, the input context here is much smaller.
With this input, we then ask the \llm to identify the final set of edit locations{, specified by} line numbers, functions, or classes. 
Our simple hierarchical localization allows \tech to select a set of relevant code snippets as edit locations for repair{}.

\subsection{Repair \protect\smugcat}

In the repair stage, the goal is to produce the correct patch to solve the issue. 
Following existing work on \llm{}-based program repair~\cite{alpharepair, kolak2022patch, codexrepair, jiang2023study}, we first utilize the identified edit locations and construct a context window of code snippets to provide to the \llm for repair.
For example, if the identified location was a class from line 40 to 78, we would produce a context window of \CodeIn{[40 - x, 78 + x]} where \CodeIn{x} denotes the context window size. 
The intuition behind adding the additional code before and after the identified location is to provide the \llm with relevant contextual information for better program repair~\cite{alpharepair}. 
If multiple edit locations are identified, we would concatenate these context windows together separated with ``\CodeIn{...}'' to indicate missing context in the middle.

Using the code snippets, we then ask the \llm to generate patches to solve the issue.  
However, instead of directly producing the entire code snippet to replace the entire given context, \tech asks the \llm to generate a \textbf{Search/Replace edit}~\cite{aidar}: a simple 
diff format to efficiently create each patch. 
Figure~\ref{fig:diff_format} shows an example of the Search/Replace format containing two main parts: 1) search: the original code snippet we want to replace and 2) replace: the replacement code snippet we want to replace with.
To apply the generated Search/Replace diff to the original file, we can simply match the search code snippet and replace it with the replacement. 
This simple diff format avoids generating the complete code and instead focuses on producing small edits, which are not only more cost-efficient, but also more reliable and accurate (less chances for hallucination).
For each issue, \tech uses the \llm to generate multiple potential patches (starting with greedy and then sample multiple patches with higher temperature).

 \subsection{Patch Validation \protect\grinningcat}

\begin{wrapfigure}{r}{0.4\textwidth}
  \begin{center}
    \includegraphics[width=0.4\textwidth]{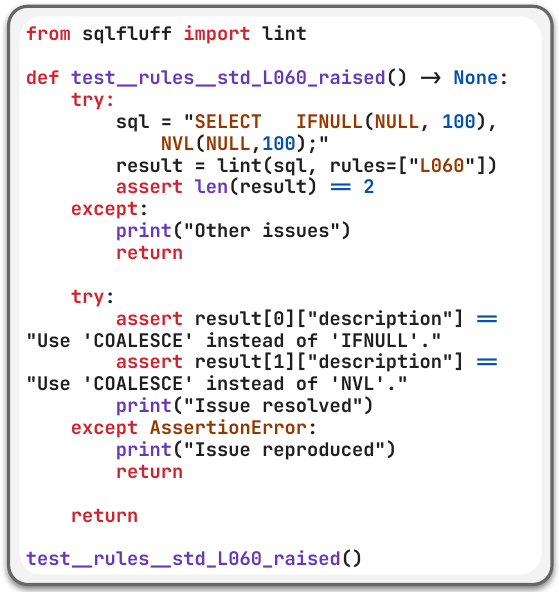}
  \end{center}
  \caption{Example \reproduction test.}
  \label{fig:test}
\end{wrapfigure}

 \subsubsection{Reproduction test generation.} 
 \label{sec:reproduction_test_generation}
 Since \tech generates multiple candidate patches per issue, we need a way to select a final patch for submission.
 Please note that under the realistic \swebench setup, the original project codebase can only provide regression tests and not any \reproduction tests (i.e., bug-triggering tests). 
 This is because the issue has just been raised and the developers have not added any additional tests to trigger the issue.  
 As such, different from the Generate-and-Validate program repair setup~\cite{long2016analysis}, we do not have access to bug-triggering tests. %

Following prior work~\cite{specrover, coder}, \tech generates additional \reproduction test to help with patch selection. More specifically, \tech leverages the \llm to synthesize a complete testing file that attempts to both reproduce the original issue described in the issue description, as well as verify whether the issue has been fixed.
 Figure~\ref{fig:test} shows an example of the \reproduction test that we want the model to synthesize. 
 {If the issue is reproduced}, the test outcome should print \CodeIn{Issue reproduced}. 
 On the other hand, the test should output \CodeIn{Issue resolved} if the issue has been fixed. 
 We also include another output of \CodeIn{Other issues} if the test runs into any unexpected issues. 
 To generate the \reproduction test, \tech provides the original issue description with an example \reproduction test to demonstrate the test format.
 Similar to repair, we also sample multiple candidate \reproduction tests and then execute each test on the original repository to filter any tests that do not output \CodeIn{Issue reproduced}. 
 Finally, we normalize each test (remove comments, extra spaces, and normalize test names) and then select the test with the highest number of occurrence as the final \reproduction test for each issue.

\subsubsection{Patch selection.} 
\label{sec:patch_filtering}

Using the generated \reproduction tests, we start our patch selection process to pick the final submission patch.
\tech first runs all the existing tests in the repository to identify a set of passing tests that successfully pass in the original codebase. 
However, not all of those passing tests should be considered as regression tests since solving the issue may require changing some of the existing functionalities. 
Therefore, \tech provides the list of passing tests to the \llm and ask it to identify any tests that should not be ran to check if the issue has been correctly fixed (\ie the tests that may be updated/patched during issue fixing). 
After removing the \llm-identified non-regression tests, we obtain a final set of regression tests. 
We then run the set of regression tests on all the generated patches. 
\tech then keeps the patches with the lowest number of regression failures.
For those patches, \tech then runs the selected \reproduction test and only keeps patches that output \CodeIn{Issue resolved}.
Meanwhile, because the \reproduction test is generated by the \llm and can potentially be incorrect/imprecise, it could be possible that no patch can pass the \reproduction test; in that case, \tech will fall back on only using the regression test results for selection.
\tech then applies a re-ranking approach using majority voting:
We first normalize each patch to ignore surface-level differences (\eg extra spaces, newlines, and comments), and then select the patch with the highest number of occurrences as the final patch for submission. 
More specifically, to standardize the patch, we begin by parsing both the old and new code (after applying the patch) into abstract syntax trees. Next, we unparse the trees into a canonical source code format with docstrings removed. Finally, we compute the textual diff between the standardized old and new code to get the normalized patch.

\tech solves repository-level issues using a simple step-by-step procedure.
We note here that none of the techniques used by \tech in isolation are{ revolutionary},
but instead \tech smartly combines existing techniques to construct an easy-to-understand approach. 
Different from prior autonomous agent-based tools that involve complex interactions with the environment, \tech uses{ a simplistic} three-phase approach to localize, repair, and validate without relying on any agents for decision-making.
By conducting localization in a hierarchical manner, \tech can efficiently and effectively compute the fine-grained locations for editing. 
\tech then performs repair by sampling multiple patches using a simple diff format.
\tech's patch validation approach can further aid the patch selection process by producing \reproduction tests that can help verify if the issue is fixed.

%% file: secs/experimental.tex
\section{Experimental Setup}

\parabf{Datasets.}
We evaluate \tech and baselines using the popular \swebench{} dataset~\cite{swebench} to test the ability to solve real-world software engineering issues. 
Each problem in \swebench requires submitting a patch to solve the underlying issue described in the input issue description. 
In particular, we focus on the widely used  \swebenchlite{} version~\cite{swebenchlite}, containing 300 self-contained problems with better quality.
Furthermore, we also conduct a detailed study (Section~\ref{sec:benchmark_analysis}) on the \swebenchlite{} benchmark to not only demonstrate potential issues and biases, but also produce a more rigorous filtered set of problems for better evaluation. %

\parabf{Implementation.}
We implement \tech using \gptfouro{} (\CodeIn{gpt-4o-2024-05-13})~\cite{gpt4o}.
By default, we query the \llm with greedy decoding. 
During sampling, we use a sampling temperature of $0.8$.
For the embedding-based retrieval method, we implement our approach using LlamaIndex~\cite{llamaindex}. 
We use OpenAI's \CodeIn{text-embedding-3-small}~\cite{openaiembedding} model to compute the embedding with chunk size of 512 and chunk overlap of 0.
For each issue, we first localize to the top three suspicious files, and then localize to an unrestricted number of suspicious classes and functions within these files, all using greedy decoding.
Next, to maximize the chances of finding the correct edit locations, we draw four samples of edit locations per issue (i.e., the third step in the localization phase).
This gives us 4 separate sets of edit locations per issue.
For each set, we adopt a context window of $\pm$ 10 lines around each edit location, and generate 10 patches (1 greedy and 9 samples).
This results in a total of 40 patches per bug.
We adopt the same Search/Replace edit format from prior work~\cite{aidar}, and use the built-in Python \CodeIn{ast} library~\cite{ast} to perform parsing in our normalization step.
To generate the \reproduction tests, we also generate 40 samples (1 greedy and 39 samples) in total prior to patch selection (described in Section~\ref{sec:reproduction_test_generation}).
The regression tests are obtained by first running all the tests to obtain a set of passing tests that successfully pass in the original repository and then use the \llm to identify any non-regression tests (described in Section~\ref{sec:patch_filtering}). 
We do not directly use the provided list of regression tests already identified in the \CodeIn{PASS\_TO\_PASS} field of \swebench as requested by the \swebench maintainers\footnote{If we directly use the \CodeIn{PASS\_TO\_PASS} tests the performance on \swebenchlite{} will be 98.}. 
We modify the official \swebench evaluation setup to be able to freely execute arbitrary regression and \reproduction tests.

\parabf{Baselines.} 
We compare \tech against \totalbaselines agent-based approaches. 
These baseline tools represent the state-of-the-art performance on \swebench{}. 
We include state-of-the-art open-source as well as commercial or closed-source baselines (indicated via a \lock{}). 
We note here that the majority of the closed-source baselines do not provide any trajectories, just the submission patches. Therefore, we cannot verify the steps taken to arrive at the final patches.
Moreover, we also include a simple agentless baseline using retrieval-augmented generation (RAG) proposed as part of \swebench~\cite{swebench} for comparison.
In this case, the agentless baseline uses the \llm to directly generate a patch file by providing it with the file content of the most relevant files, retrieved using BM25~\cite{robertson2009probabilistic}. 

\parabf{Metrics.}
Following prior work~\cite{autocoderover}, we report \textbf{1) \% Resolved}: the percentage of resolved problems in the benchmark, \textbf{2) Avg. \$ Cost}: average inference cost of running the tool, and \textbf{3) Avg. \# Tokens}: average number of input and output tokens used to query to \llm. 
Additionally, we also report the \textbf{\% Correct Location}: the percent of problems where the patch produced by the tool covers the edit location(s) of the ground truth developer patch.
We compute this metric over three granularities: file, function, and line. 
We report that a patch contains the correct location if it edits a superset of all locations in the ground truth patch.
For baseline tools, we directly use the reported results either from the official leaderboard~\cite{swebenchleaderboard} or from the tool's official paper/repository. 

%% file: secs/eval.tex
\section{Evaluation}

\begin{table}[h]
\centering
\caption{Results on \swebenchlite{}. Note: \protect\lock{} indicates approaches that are closed-source (\ie source code is not released).
\protect\cat{} \textbf{indicates approaches built on top of an earlier version of our \tech}.
\textquotesingle-\textquotesingle{} indicates that the relevant information to compute this has not been released. \protect\questionmark{} indicates that multiple models are used, but some of them are not specified. 
\claudesonnet{} is short for \claudesonnet{}onnet.}

\label{tab:main}
\scalebox{0.70}{
\begin{tabular}{ll|rrrrrr}
\toprule
 \multirow{2}*{Tool} & \multirow{2}*{\llm} & \multirow{2}*{\%{} Resolved} & \multirow{2}*{\makecell{Avg.\\{}\$ Cost}} & \multirow{2}*{\makecell{Avg.\\{}\# Tokens}} & \multicolumn{3}{c}{\% Correct Location} \\
 & & & & & Line & Function & File \\
\midrule

\cellcolor[HTML]{e1dcef}\codestoryaide{}~\cite{codestoryaide} \lock{} & \cellcolor[HTML]{e1dcef}\makecell[l]{\openai{} \gptfouro{}+\anthropic{} \claudesonnet} & \cellcolor[HTML]{e1dcef}129 (43.00\%) & \cellcolor[HTML]{e1dcef} - &\cellcolor[HTML]{e1dcef} - &\cellcolor[HTML]{e1dcef}41.7\% &\cellcolor[HTML]{e1dcef}58.7\% &\cellcolor[HTML]{e1dcef}72.0\% \\
\bytedanceagent{}~\cite{marscode} \lock{} & NA & 118 (39.33\%) &  - & - &42.7\% &58.0\% &79.7\% \\
\cellcolor[HTML]{e1dcef}\honeycomb{}~\cite{honeycomb} \lock{} & \cellcolor[HTML]{e1dcef}NA & \cellcolor[HTML]{e1dcef}115 (38.33\%) & \cellcolor[HTML]{e1dcef} - &\cellcolor[HTML]{e1dcef} - &\cellcolor[HTML]{e1dcef}44.3\% &\cellcolor[HTML]{e1dcef}57.0\% &\cellcolor[HTML]{e1dcef}69.3\% \\
\mentatbot{}~\cite{mentatbot} \lock{} & \openai{} \gptfouro{} & 114 (38.00\%) &  - & - &37.3\% &53.3\% &69.3\% \\
\cellcolor[HTML]{e1dcef}\gru{}~\cite{gru} \lock{} & \cellcolor[HTML]{e1dcef}NA & \cellcolor[HTML]{e1dcef}107 (35.67\%) & \cellcolor[HTML]{e1dcef} - &\cellcolor[HTML]{e1dcef} - &\cellcolor[HTML]{e1dcef}38.3\% &\cellcolor[HTML]{e1dcef}54.3\% &\cellcolor[HTML]{e1dcef}75.0\% \\
\isoform{}~\cite{isoform} \lock{} & NA & 105 (35.00\%) &  - & 41,963 &38.7\% &55.3\% &72.0\% \\
\cellcolor[HTML]{e1dcef}\supercoder{}~\cite{supercoder} \lock{} & \cellcolor[HTML]{e1dcef}NA & \cellcolor[HTML]{e1dcef}102 (34.00\%) & \cellcolor[HTML]{e1dcef} - &\cellcolor[HTML]{e1dcef} - &\cellcolor[HTML]{e1dcef}41.7\% &\cellcolor[HTML]{e1dcef}63.7\% &\cellcolor[HTML]{e1dcef}65.7\% \\
\lingma{}~\cite{lingma} \lock{} & \makecell[l]{\openai{} \gptfouro{}+\anthropic{} \claudesonnet} & 99 (33.00\%) &  - & - &40.0\% &58.7\% &75.0\% \\
\cellcolor[HTML]{e1dcef}\factorydroid{}~\cite{factorydroid} \lock{} & \cellcolor[HTML]{e1dcef}NA & \cellcolor[HTML]{e1dcef}94 (31.33\%) & \cellcolor[HTML]{e1dcef} - &\cellcolor[HTML]{e1dcef} - &\cellcolor[HTML]{e1dcef}36.7\% &\cellcolor[HTML]{e1dcef}55.7\% &\cellcolor[HTML]{e1dcef}72.7\% \\
\amazonqagent{}-v2~\cite{amazonqdeveloper} \lock{} & NA & 89 (29.67\%) &  - & - &40.3\% &52.0\% &74.3\% \\
\cellcolor[HTML]{e1dcef}\specrover{}~\cite{specrover} \lock{} & \cellcolor[HTML]{e1dcef}\makecell[l]{\openai{} \gptfouro{}+\anthropic{} \claudesonnet} & \cellcolor[HTML]{e1dcef}93 (31.00\%) & \cellcolor[HTML]{e1dcef}\$0.65 &\cellcolor[HTML]{e1dcef} - &\cellcolor[HTML]{e1dcef} - &\cellcolor[HTML]{e1dcef} - &\cellcolor[HTML]{e1dcef} - \\
\coder{}~\cite{coder} \lock{} & \openai{} \gptfour{} & 85 (28.33\%) & \$3.34 & 323,802 &35.7\% &52.3\% &67.0\% \\
\cellcolor[HTML]{e1dcef}\masai{}~\cite{masai} \lock{} & \cellcolor[HTML]{e1dcef}NA & \cellcolor[HTML]{e1dcef}84 (28.00\%) & \cellcolor[HTML]{e1dcef} - &\cellcolor[HTML]{e1dcef} - &\cellcolor[HTML]{e1dcef}38.7\% &\cellcolor[HTML]{e1dcef}56.3\% &\cellcolor[HTML]{e1dcef}75.0\% \\
\sima{}~\cite{sima} \lock{} & \openai{} \gptfouro{} & 83 (27.67\%) & \$0.82 & - &37.0\% &54.0\% &79.0\% \\
\cellcolor[HTML]{e1dcef}\ibmagent{}~\cite{ibmagent} \lock{} & \cellcolor[HTML]{e1dcef}NA & \cellcolor[HTML]{e1dcef}80 (26.67\%) & \cellcolor[HTML]{e1dcef} - &\cellcolor[HTML]{e1dcef} - &\cellcolor[HTML]{e1dcef}39.7\% &\cellcolor[HTML]{e1dcef}56.7\% &\cellcolor[HTML]{e1dcef}73.3\% \\
\opencsgagent{}~\cite{opencsgstarship} \lock{} & \openai{} \gptfour{} & 71 (23.67\%) &  - & - &39.0\% &61.7\% &90.7\% \\
\cellcolor[HTML]{e1dcef}\amazonqagent{}~\cite{amazonqdeveloper} \lock{} & \cellcolor[HTML]{e1dcef}NA & \cellcolor[HTML]{e1dcef}61 (20.33\%) & \cellcolor[HTML]{e1dcef} - &\cellcolor[HTML]{e1dcef} - &\cellcolor[HTML]{e1dcef}34.0\% &\cellcolor[HTML]{e1dcef}43.7\% &\cellcolor[HTML]{e1dcef}71.7\% \\
\repounderstander{}~\cite{repounderstander} \lock{} & \openai{} \gptfour{} & 64 (21.33\%) &  - & - & - & - & - \\
\midrule
\cellcolor[HTML]{e1dcef}\autocoderover{}-v2~\cite{autocoderovertwo} & \cellcolor[HTML]{e1dcef}\openai{} \gptfouro{} & \cellcolor[HTML]{e1dcef}92 (30.67\%) & \cellcolor[HTML]{e1dcef} - &\cellcolor[HTML]{e1dcef} - &\cellcolor[HTML]{e1dcef}35.0\% &\cellcolor[HTML]{e1dcef}52.3\% &\cellcolor[HTML]{e1dcef}69.3\% \\
\repograph~\cite{repograph} & \openai{} \gptfouro{} & 89 (29.67\%) &  - & - &36.7\% &51.3\% &71.0\% \\
\cellcolor[HTML]{e1dcef}\multirow{1}*{\moatless{}~\cite{moatless}} & \cellcolor[HTML]{e1dcef}\anthropic{} \claudesonnet{} & \cellcolor[HTML]{e1dcef}80 (26.67\%) & \cellcolor[HTML]{e1dcef}\$0.17 &\cellcolor[HTML]{e1dcef} - &\cellcolor[HTML]{e1dcef}38.7\% &\cellcolor[HTML]{e1dcef}54.7\% &\cellcolor[HTML]{e1dcef}78.7\% \\
\cellcolor[HTML]{e1dcef} & \cellcolor[HTML]{e1dcef}\openai{} \gptfouro{} & \cellcolor[HTML]{e1dcef}74 (24.67\%) & \cellcolor[HTML]{e1dcef}\$0.14 &\cellcolor[HTML]{e1dcef} - &\cellcolor[HTML]{e1dcef}36.0\% &\cellcolor[HTML]{e1dcef}52.0\% &\cellcolor[HTML]{e1dcef}73.0\% \\
\opendevincodeact{}~\cite{opendevin} & \anthropic{} \claudesonnet & 80 (26.67\%) & \$1.14 & - &38.0\% &49.7\% &67.3\% \\
\cellcolor[HTML]{e1dcef}\aider{}~\cite{aidar} & \cellcolor[HTML]{e1dcef}\makecell[l]{\openai{} \gptfouro{}+\anthropic{} \claudesonnet} & \cellcolor[HTML]{e1dcef}79 (26.33\%) & \cellcolor[HTML]{e1dcef} - &\cellcolor[HTML]{e1dcef} - &\cellcolor[HTML]{e1dcef}35.3\% &\cellcolor[HTML]{e1dcef}50.0\% &\cellcolor[HTML]{e1dcef}69.7\% \\
\multirow{1}*{\sweagent{}~\cite{sweagent}} & \anthropic{} \claudesonnet{} & 69 (23.00\%) & \$1.62 & 521,208 &40.7\% &54.3\% &72.0\% \\
 & \openai{} \gptfouro{} & 55 (18.33\%) & \$2.53 & 498,346 &29.3\% &42.3\% &58.3\% \\
 & \openai{} \gptfour{} & 54 (18.00\%) & \$2.51 & 245,008 &30.7\% &45.3\% &61.0\% \\
\cellcolor[HTML]{e1dcef}\appmapnavie{}~\cite{appmapnavie} & \cellcolor[HTML]{e1dcef}\openai{} \gptfouro{} & \cellcolor[HTML]{e1dcef}65 (21.67\%) & \cellcolor[HTML]{e1dcef} - &\cellcolor[HTML]{e1dcef} - &\cellcolor[HTML]{e1dcef}29.7\% &\cellcolor[HTML]{e1dcef}44.7\% &\cellcolor[HTML]{e1dcef}59.7\% \\
\autocoderover{}~\cite{autocoderover} & \openai{} \gptfour{} & 57 (19.00\%) & \$0.45 & 38,663 &29.0\% &42.3\% &62.3\% \\
\midrule
\cellcolor[HTML]{e1dcef}\rag{}~\cite{sweagent} & \cellcolor[HTML]{e1dcef}\anthropic{} \claudeopus{} & \cellcolor[HTML]{e1dcef}13 (4.33\%) & \cellcolor[HTML]{e1dcef}\$0.25 &\cellcolor[HTML]{e1dcef} - &\cellcolor[HTML]{e1dcef}22.0\% &\cellcolor[HTML]{e1dcef}30.0\% &\cellcolor[HTML]{e1dcef}57.0\% \\
\cellcolor[HTML]{e1dcef} & \cellcolor[HTML]{e1dcef}\openai{} \gptfour{} & \cellcolor[HTML]{e1dcef}8 (2.67\%) & \cellcolor[HTML]{e1dcef}\$0.13 &\cellcolor[HTML]{e1dcef} - &\cellcolor[HTML]{e1dcef}12.7\% &\cellcolor[HTML]{e1dcef}23.3\% &\cellcolor[HTML]{e1dcef}47.3\% \\
\cellcolor[HTML]{e1dcef} & \cellcolor[HTML]{e1dcef}\anthropic{} \claude{}-2 & \cellcolor[HTML]{e1dcef}9 (3.00\%) & \cellcolor[HTML]{e1dcef} - &\cellcolor[HTML]{e1dcef} - &\cellcolor[HTML]{e1dcef}16.7\% &\cellcolor[HTML]{e1dcef}24.3\% &\cellcolor[HTML]{e1dcef}46.7\% \\
\cellcolor[HTML]{e1dcef} & \cellcolor[HTML]{e1dcef}\openai{} \chatgpt{} & \cellcolor[HTML]{e1dcef}1 (0.33\%) & \cellcolor[HTML]{e1dcef} - &\cellcolor[HTML]{e1dcef} - &\cellcolor[HTML]{e1dcef}6.3\% &\cellcolor[HTML]{e1dcef}11.3\% &\cellcolor[HTML]{e1dcef}27.3\% \\
\midrule
\textbf{\tech{}} & \openai{} \gptfouro{} & 96 (32.00\%) & \$0.70 & 78,166 &35.3\% &52.0\% &69.7\% \\

\bottomrule
\end{tabular}
}
\end{table}

\subsection{Performance on \swebenchlite{}}

Table~\ref{tab:main} shows the main evaluation result of \tech and prior agent-based approaches on \swebenchlite{}. 
We observe that \tech is able to solve \totalsolve out of 300 problems (\totalsolveperc{}).  
While this is not the highest percentage of problems solved on \swebenchlite{}, \tech is extremely competitive compared with prior agent-based approaches while using a much simpler design and overall technique. 
It is important to note here that many of the top techniques are closed-source/commercial and did not release any source code to reproduce experiments or even {trajectories} for further verification. 
\textbf{Compared with all open-source approaches, \tech is able to achieve the highest performance of \totalsolveperc{} (\totalsolve{} / 300) on \swebenchlite.}
Additionally, \tech only costs on average \averagedollarcost{}, which is less than most prior agent-based approaches. 
Comparing against the RAG agentless baselines, we see that while \tech costs slightly more, \tech is also able to fix way more issues.

\begin{wrapfigure}{r}{0.55\textwidth}
  \begin{center}
    \includegraphics[width=0.55\textwidth]{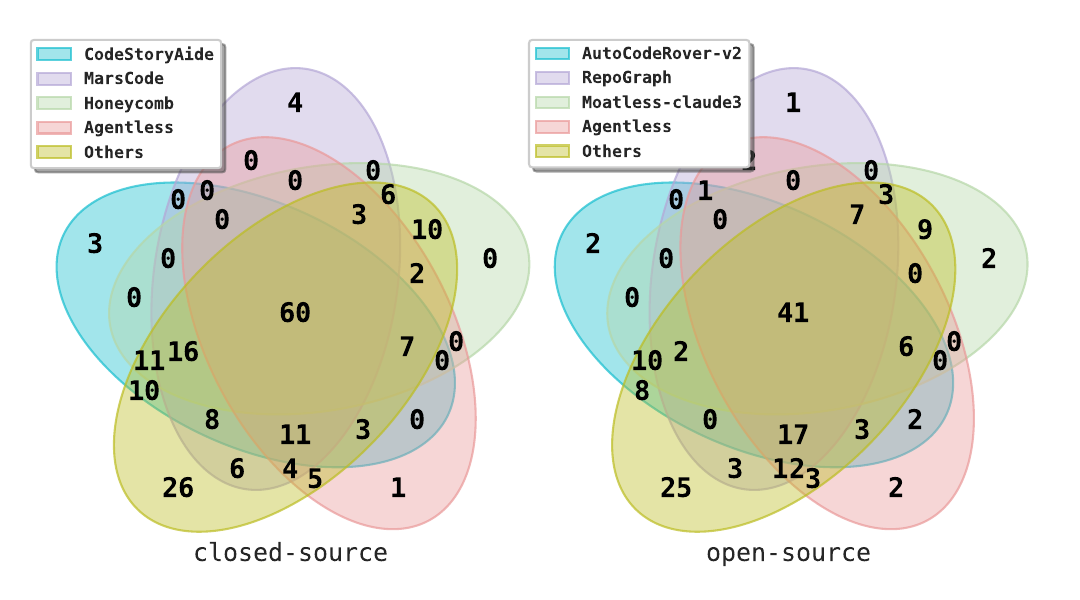}
  \end{center}
  \caption{Venn diagram for issue fixes.}
  \label{fig:venn}
\end{wrapfigure}

\subsubsection{Unique issues fixed.}  Figure~\ref{fig:venn} shows the unique issues solved by \tech compared with the top-performing closed-source / commercial and open-source approaches (``Others'' indicates all other approaches within each category). 
First, we see that compared to the open-source agent-based tools, \tech is able to {fix 2 issues that no other tools can resolve}, showing the success of using a simple agentless approach in solving difficult issues.
Furthermore, even when compared with high-performing commercial approaches, \tech is still able to offer a unique fix. 
This low number of unique fixes can be attributed to the fact there are already tools built on top of earlier versions of \tech (e.g., \isoform{}~\cite{isoform}) or partly inspired by \tech (e.g., \bytedanceagent~\cite{marscode}), thereby reducing the unique issues resolved by \tech.
Nevertheless, our results show that \tech can be competitive/complementary compared to existing agents.

\subsubsection{Localization performance.} In real-world software development, apart from directly fixing the issue, providing the correct edit location to developers is extremely helpful for debugging. 
As such, we examine the locations of the patches generated by each technique compared with the ground truth patch. 
We note here that it is possible to fix a bug in a different location than the ground truth, however comparing against the ground truth patch can still serve as an approximate measure.  
Table~\ref{tab:main} additionally shows the {percentage} of {submitted patches with correct locations} for each tool{, across} line, function, and file levels.
We first observe that the percentage of {patches with correct locations} correlates heavily with the solve rate. 
Interestingly, the highest result for file-level location is \opencsgagent{} at 90.0\%, significantly higher than even the best-performing approaches while at the same time having a relatively low solve rate (23.67\%). 
As \opencsgagent{} is a commercial product that does not provide source code or detailed trajectories, it is difficult to explain this huge difference between localization and repair performance. 
In terms of localization performance, by using our simple hierarchical approach, \tech remains very competitive compared with previous agent-based approaches.

\subsubsection{Reproduction test results.}

\tech uses regression and generated \reproduction tests to perform filtering in order to select the final submission patch. 
Therefore, we evaluate the quality of our generated \reproduction tests. 
We note here that as described in Section~\ref{sec:reproduction_test_generation}, we only use the generated \reproduction test if it can successfully reproduce the original issue in the original repository. 
Out of the 300 problems in \swebenchlite, \tech is able to produce 213 \reproduction tests that output the required reproduction message when evaluated on the original repository.
However, these tests might still be incorrect as a correct \reproduction should also be able to verify that the issue has been correctly resolved.
To evaluate this, we directly apply the ground truth patch provided in \swebench\footnote{Note that the ground truth patch is only applied here to evaluate the test quality, and is not used during the \tech process for selecting the reproduction tests.}.
We found that only 94 tests correctly output the \CodeIn{Issues resolved} message after applying the ground truth patches. 
This steep drop-off can be partially explained as sometimes the issue description provided in the problem may not contain enough information to generate complete test cases for validating a correct solution.
However, this problem is partially mitigated as \tech takes a conservative approach of requiring all patches to pass the regression test suite first and will remove a patch if it cannot pass the regression tests but passes on the generated \reproduction test. 
This reduces the likelihood of an incorrect \reproduction test selecting a correct patch. 
Furthermore, if all generated patches cannot pass the \reproduction test, \tech will fall back on only using the regression test results for selection.
In Section~\ref{sec:test_generation_ablation} we closely examine the impact of using both regression and \reproduction test for patch selection has on the performance.

\begin{wraptable}{r}{0.55\textwidth}
\footnotesize
\centering
\caption{Performance of different localization steps.}
\label{tab:ablationlocation}
\scalebox{0.85}{
\begin{tabular}{l|rrr}
\toprule
Method & Contains GT & LoC & Avg. \$\\
\midrule 

\multicolumn{4}{c}{File level localization} \\ 
\midrule

\cellcolor[HTML]{e1ecff}Prompting-based & \cellcolor[HTML]{e1ecff}78.67\% & \cellcolor[HTML]{e1ecff}3,221 & \cellcolor[HTML]{e1ecff}\$0.02 \\
\makecell[l]{Embedding-based \\ (w/o irrelevant filtering)} & 67.67\% & 3,388 & \$0.05 \\
\cellcolor[HTML]{e1ecff}\makecell[l]{Embedding-based \\ (w/ irrelevant filtering)}& \cellcolor[HTML]{e1ecff}70.33\% & \cellcolor[HTML]{e1ecff}3,622 & \cellcolor[HTML]{e1ecff}\$0.04 \\
\textbf{Combined} & 81.67\% & 3,424 & \$0.06 \\
\midrule
\multicolumn{4}{c}{Related element localization} \\
\midrule
\cellcolor[HTML]{e1ecff}Complete file & \cellcolor[HTML]{e1ecff}53.67\% & \cellcolor[HTML]{e1ecff}778 & \cellcolor[HTML]{e1ecff}\$0.15 \\
\textbf{Skeleton format} & 58.33\% & 698 & \$0.02 \\
\midrule
\multicolumn{4}{c}{Edit location localization} \\
\midrule
\cellcolor[HTML]{e1ecff}Greedy & \cellcolor[HTML]{e1ecff}50.67\% & \cellcolor[HTML]{e1ecff}189 & \cellcolor[HTML]{e1ecff}\$0.06 \\
Direct from file-level & 47.00\% & 208 & \$0.18 \\
\cellcolor[HTML]{e1ecff}Multi-samples merged & \cellcolor[HTML]{e1ecff}56.33\% & \cellcolor[HTML]{e1ecff}342 & \cellcolor[HTML]{e1ecff}\$0.07 \\
\multirow{4}*{\textbf{Multi-samples}} & 49.67\% & 165 & \multirow{4}*{\$0.07} \\
 & 49.33\% & 180 &  \\
 & 49.33\% & 168 &  \\
 & 48.33\% & 213 &  \\

\bottomrule
\end{tabular}
}
\end{wraptable}

\subsubsection{Adoption of \tech.} 
Although released very recently, \tech has already received widespread adoption. 
In Table~\ref{tab:main}, there are two baseline tools (indicated via \cat) already built upon earlier versions of our \tech:
\repograph~\cite{repograph} is an open-source tool combining \tech with repository-level graph, while \isoform{}~\cite{isoform} is a closed-source commercial tool also built upon \tech. 
Additionally, \bytedanceagent~\cite{marscode} is partly inspired by \tech in patch sampling/selection and OpenDevin~\cite{opendevin} is in the process of integrating \tech into their ecosystem. 
Furthermore, \tech has also been adopted by OpenAI as the default approach when showcasing the real-world coding performance of GPT-4o on their new \swebenchverified benchmark~\cite{openaiverified}, where \tech also achieves the best performance compared with all other studied agent-based solutions.
At the time of finishing up this draft, OpenAI just released the new o1 model family and also adopted \tech as the top/default approach to showcase their performance on \swebench~\cite{openaioonesystemcard}.%

\subsection{Ablation study on components of \tech} 

Next, we look at how each component in the localization, repair, and patch validation phases contributed to the final \tech performance. 
Unless otherwise specified, we vary the configuration of one component while using the default parameters for all other settings. 

\subsubsection{Localization ablation.}

Table~\ref{tab:ablationlocation} shows the performance and cost for each of the 3 steps in \tech's localization phase.
We show after each localization step the percentage of problems whose ground truth edit locations remain in the location set (``Contains GT''), the average lines of code {of each location set (``LoC'')}, and the average dollar cost of each step (``Avg.\$'').
The \textbf{bold} method indicates the default setting of \tech. 
First, we examine the different configurations of file level localization. 
To start with, for the retrieval method based on embeddings, we see that without including the irrelevant folder filtering to remove irrelevant folders for embedding (described in Section~\ref{sec:localize_suspicious_files}), both the performance and cost become worse. 
This demonstrates the importance of limiting the number of files to consider during embedding and focusing on essential parts of the repository for more cost-efficient and effective localization.
We see that using the prompting-based or the embedding-based retrieval method alone can locate the ground truth file in 78.7\% and 67.7\% of cases respectively. 
This can be further improved by combining them to obtain 81.7\% correct file localization, showing that prompting-based and embedding-based retrieval methods can complement each other in identifying different sets of {relevant} files.

\begin{wraptable}{r}{0.55\textwidth}
\footnotesize
\centering
\caption{Performance of different repair setups.}
\label{tab:ablationrepair}
\scalebox{0.9}{
\begin{tabular}{l|rr}
\toprule
Method & Performance & Avg. \$ \\
\midrule

\cellcolor[HTML]{e1dcef}\makecell[l]{Greedy location\\(40 samples)} & \cellcolor[HTML]{e1dcef}88 (29.33\%) & \cellcolor[HTML]{e1dcef}\$0.22 \\
\makecell[l]{Multi-samples merged \\(40 samples)} & 85 (28.33\%) & \$0.24 \\
\cellcolor[HTML]{e1dcef}\makecell[l]{\textbf{Multi-samples}\\{\textbf{(4 x 10 samples)}}} & \cellcolor[HTML]{e1dcef}96 (32.00\%) & \cellcolor[HTML]{e1dcef}\$0.29 \\

\bottomrule
\end{tabular}
}
\end{wraptable}

Using all of the localized files leads to a large context window ($>$3000).
As such, in our second localization step, we localize to the relevant classes and functions, and are able to drastically reduce the context window ($<$800). 
We compare our input of using skeleton format (described in Section~\ref{sec:localize_related_elements}) to provide a more concise representation with the baseline of using the complete file content.
We observe that by using the complete file content, not only is the cost much higher but also the number of localized groundtruth issues is reduced. 
The reason is most likely that \llm{s} cannot handle long context very well, so providing the entire file contents can confuse the model.
Conversely, by using a more concise representation as the input, we can effectively localize the correct related locations that are needed for inspection and editing.

Next, \tech localizes to the exact edit locations needed to achieve even more context reduction without losing much of the localization accuracy. 
We compare the different ways we can perform the edit location localization: 1) ``Greedy'': using greedy decoding to obtain one set of edit locations, 2) ``Direct from file-level'': directly go from file-level localization to the edit locations (instead the default of localizing from file-level to related elements and then to edit locations), 3) ``Multi-samples merged'': sample multiple sets of edit locations and merging them into one set, and 4) ``Multi-samples'': sample multiple sets of edit locations. 
We first observe that by directly going from file-level to the edit locations, both the cost and performance are worse. 
The reason is that the model can become confused when providing a large context, demonstrating the importance of our hierarchical localization design.
We also find that when merging multiple samples together, the amount of ground truth localized is higher but at the expense of having to add more context as the input during the repair phase.
Our default settings also sample the edit locations multiple times, however instead of merging, we perform repair on them separately to make use of the fact each sampled location set can provide similar localization performance while also limiting the input context.
In short, by using hierarchical localization steps, \tech can successfully minimize the cost while performing effective localization.

\subsubsection{Repair ablation.}
We now look at the impact of our different repair setups on the final performance. Table~\ref{tab:ablationrepair} shows the different settings and inputs for our repair phase with their performance and cost.%
Starting with using the greedy location set generated in the edit location generation stage, we observe that we can already achieve very high performance of more than 88 issues fixed.
Similarly, for the ``Multi-samples merged'' where we merged multiple location sets into one, we can also achieve comparable performance.
The performance can be further improved by considering each sampled locations separately (to generate 10 candidate patches each) when performing repair to achieve \totalsolve fixes.
The reason is that each different location sets may localize different ground truth locations and provide different context that can be helpful to fix specific issues.
By using different edit locations and combining with our extensive test filtering and selection stage, \tech can drastically improve the repair performance.

\begin{wrapfigure}{r}{0.4\textwidth}
  \begin{center}
    \includegraphics[width=0.4\textwidth]{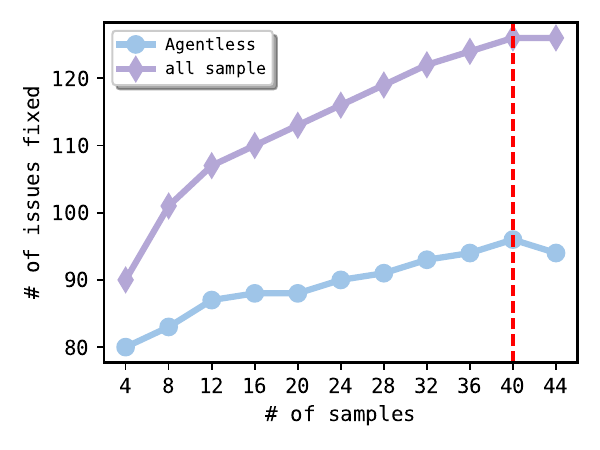}
  \end{center}
  \caption{Repair performance as the number of patch samples increases.}
  \label{fig:repair_performance_samples}
\end{wrapfigure}

Next, we examine the impact of using different numbers of sampled candidate patches on the performance of \tech.
Figure~\ref{fig:repair_performance_samples} shows the number of issues fixed as we increase the number of samples. 
Note that the sample interval increases by 4 since we use 4 different sets of locations as input. 
First, we see that by just using 1 greedy sample for each location set, \tech can already achieve a significant number of correct fixes of 80.
We can continue to improve repair performance by adding more samples.
However, we observe that the performance plateaus at around 40 samples where adding additional candidate patches does not improve performance. 
This is because we perform majority voting after test filtering to select the final submission patch, which means that later samples may be ignored since it is difficult for them to offset the majoritively voted patch.  
Interestingly, we can also see from the figure that \emph{if we consider all patch samples (instead of only selecting one patch) for each issue, the total number of possible issues that \tech can solve is 126 (42.0\%).} 
This shows a high upper bound for the potential of \tech with future work being even better patch re-ranking and selection techniques to further improve the overall performance.

\subsubsection{Patch validation ablation.} 
\label{sec:test_generation_ablation}

\begin{wraptable}{r}{0.45\textwidth}
\footnotesize
\centering
\caption{Performance of different patch selection.}
\label{tab:ablationselection}
\scalebox{0.9}{
\begin{tabular}{l|rr}
\toprule
Method & Performance & Avg. \$ \\
\midrule

\cellcolor[HTML]{d9ead3}Majority voting & \cellcolor[HTML]{d9ead3}77 (25.67\%) & \cellcolor[HTML]{d9ead3}\$0.00 \\
+Regression test & 81 (27.00\%) & \$0.01 \\
\cellcolor[HTML]{d9ead3}\textbf{+Reproduction test} & \cellcolor[HTML]{d9ead3}96 (32.00\%) & \cellcolor[HTML]{d9ead3}\$0.25 \\

\bottomrule
\end{tabular}
}
\end{wraptable}

Finally, we examine the impact of our different test generation and patch selection configurations has on performance.
Table~\ref{tab:ablationselection} shows the result and additional cost of different approaches. 
We see that by only using majority voting, we can already achieve 77 correct fixes.
{
By adding the existing regression tests, and filter for candidate patches with the lowest amount of regression errors, we can improve performance to 81 issues resolved.}
Furthermore, the most significant performance improvement was achieved by incorporating additional filtering based on the generated reproduction tests, resulting in the final \tech performance of \totalsolve fixes. 
This demonstrates the impact of our patch selection approach, specifically our \reproduction test generation, which is able to make use of the high number of candidate patches generated and filter for the correct patch for final submission.
However, using \reproduction tests also comes with additional costs as \tech needs to generate these tests which are not provided in the original project repository.

\begin{wrapfigure}{r}{0.4\textwidth}
  \begin{center}
    \includegraphics[width=0.4\textwidth]{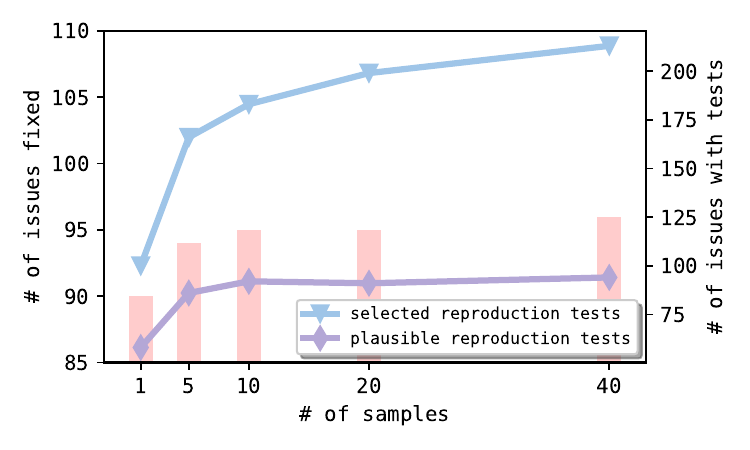}
  \end{center}
  \caption{Reproduction test and repair results as number of samples increases.}
  \label{fig:reproduction_test_samples}
\end{wrapfigure}

Next, we look at our \reproduction test generation strategy. 
Figure~\ref{fig:reproduction_test_samples} shows the repair performance (bar, left axis), the total number of selected and \plausiblecorrect \reproduction tests (lines, right axis) as we increase the number of candidate \reproduction tests generated per issue. 
Recall that we only select the \reproduction test if it can successfully output the \CodeIn{Issue reproduced} message when evaluated on the original repository.
We further consider a selected \reproduction test as \plausiblecorrect if it can successfully verify that the ground truth developer patch has fixed the original issue (i.e., output \CodeIn{Issue resolved}).
To begin with, when we only generate one \reproduction test (i.e., using the greedy output) per issue, we only produce 100 selected tests for all 300 issues in \swebenchlite with the repair performance being 90. 
We can increase the number of selected \reproduction tests by increasing the number of candidate reproduction tests we generate. 
However, it is interesting to note that the number of \plausiblecorrect \reproduction tests does not increase drastically (apart from going from 1 to 5 samples). 
The reason again is due to some ambiguity in the issue description where it may not contain sufficient information to reproduce and verify the issue has been resolved.
Nevertheless, we observe a small improvement in the number of issues resolved, reaching final performance of 96 correct fixes as we increase the number of candidate \reproduction tests to 40 per issue. 

%% file: secs/study.tex
\section{Additional Analysis on \swebenchlite}

\subsection{Problem Classification}
\label{sec:benchmark_analysis}

\begin{figure*}[h]
\centering
    \begin{subfigure}[b]{0.33\textwidth}
        \centering
        \includegraphics[width=\textwidth]{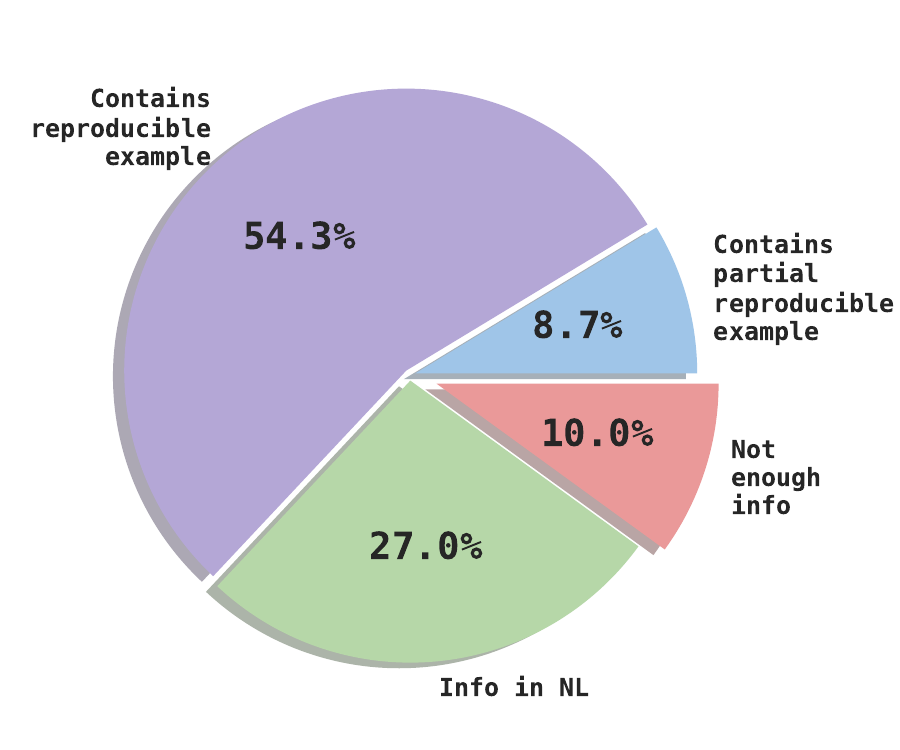}
        \caption{Description quality}
        \label{fig:benchmark_description}
    \end{subfigure}
    \begin{subfigure}[b]{0.33\textwidth}
        \centering
        \includegraphics[width=\textwidth]{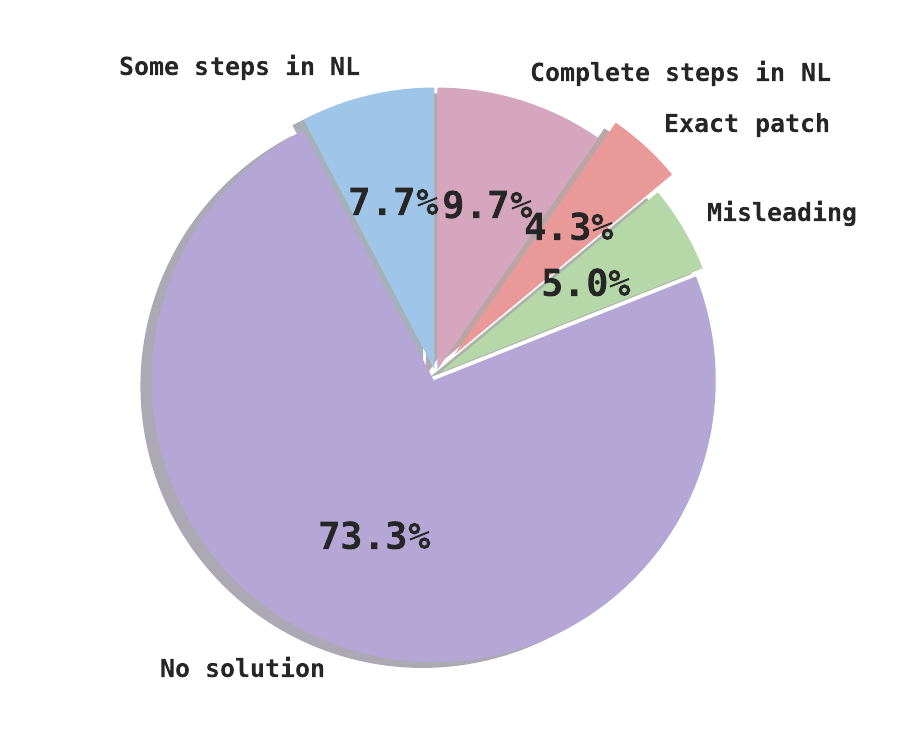}
        \caption{Solution in description}
        \label{fig:benchmark_patch}
    \end{subfigure}
    \begin{subfigure}[b]{0.32\textwidth}
        \centering
        \includegraphics[width=\textwidth]{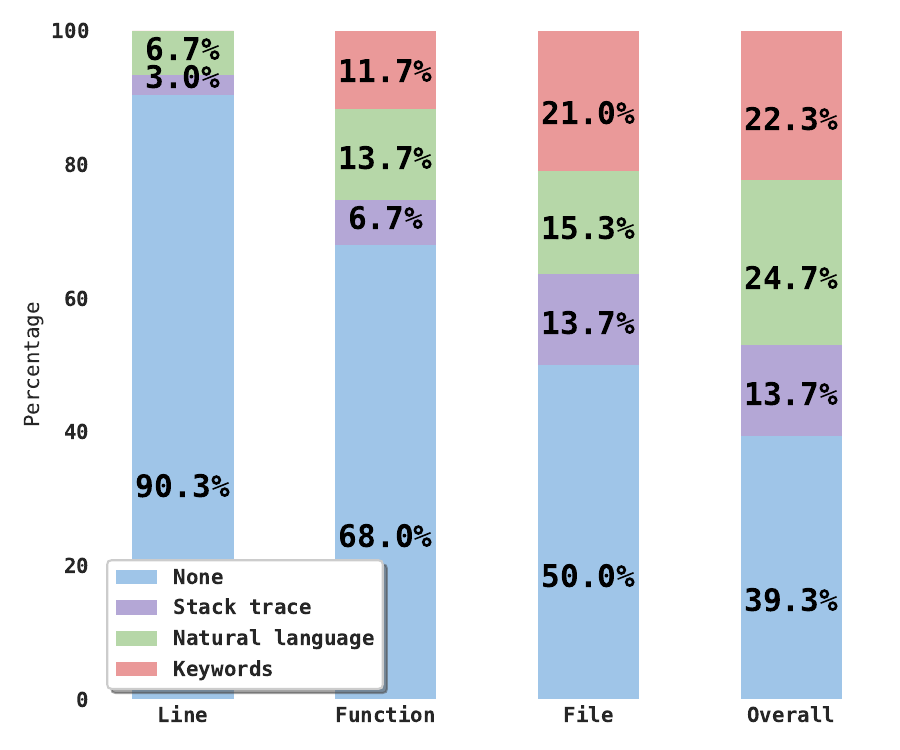}
        \caption{Location information}
        \label{fig:benchmark_location}
    \end{subfigure}
    \caption{Categorization and corresponding breakdown of the \swebenchlite problems.} 
\end{figure*}

We now take a closer look at the problems in \swebenchlite. 
We first classify the existing problems to gain better understanding and additional insights on exactly what \emph{types} of problems \tech and prior approaches can solve. 
Specifically, we perform manual classification based on the issue description and ground truth developer patch of each problem.
Below describes each of classification dimensions and their categories in more detail:

\parabf{1) Description quality.} 
We first inspect whether each issue description contains sufficient information to perform the desired task. 
Figure~\ref{fig:benchmark_description} shows the distribution of each category: \emph{(i)} contains enough information in natural language, \emph{(ii)} contains reproducible failure example, \emph{(iii)} contains partially reproducible example, and \emph{(iv)} does not contain enough information.%
We observe that while a majority of the tasks in \swebenchlite contains  sufficient information, with many having some small failure examples to showcase the bug, there is a non-trivial percentage (\notenoughinfo) of problems which do not contain enough information.
Such problems include those that require implementing a new function with a specific name or adding an error message with a specific string that was not provided in the problem description.\footnote{These types of problems still exist in the benchmark despite claims that they have been completely removed by the filtering process according to \href{www.swebench.com/lite.html}{\swebenchlite}.}
This means the test will fail if the function name or error message string does not match exactly, even if the underlying functionality is correctly implemented.
Another example of insufficient information are problems that have multiple different interpretations on how to solve the issue, and only a subset of them can pass the ground truth test.
For instance, the issue description will outline two possible solutions suggestions with only one of them aligned well with developer intention.  
Implementing the other proposed solution suggestion will lead to test failure.
This highlights the necessity to further sanitize/improve \swebenchlite{} where these problems with uninformative descriptions shall be further excluded.

\parabf{2) Solution in description.} 
We also examine whether the solution or steps to solve the problem are already provided in the issue description.
Figure~\ref{fig:benchmark_patch} shows the breakdown of our categories: {\emph{(i)} no solution or steps provided, \emph{(ii)} partial solution provided (\eg some steps in natural language), \emph{(iii)} complete solution provided (\eg complete steps in natural language), \emph{(iv)} exact patch provided, and \emph{(v)} misleading solution or steps.} 
Interestingly, we observe that \exactpatch of issues contain the exact ground truth patch in the issue description, while an additional \stepsinnl of issues describe the exact steps required to come up with the correct solution. 
This shows that certain problems in \swebenchlite can be much easier to solve since they provide the solution either in exact code snippets or natural language. 
Furthermore, we also observe \misleadingpatch of issues contain proposed solution or steps in the issue description that do not reflect the ground truth patch introduced by the developers.
This further highlights potential issues with the benchmark, as these discrepancies can mislead tools to generate incorrect solutions simply by following the issue description.

\parabf{3) Location information.} 
We further check if the issues description contains the correct location information. 
We divide the granularity into line, function, and file level. 
Our categories are: \emph{(i)} exact locations in natural language, \emph{(ii)} exact locations provided in failure stack traces, \emph{iii)} related keywords in the issue description that can be used to search for the location, and \emph{(iv)} not provided. 
We first observe that only in very few cases ($<$10\%), the issue provides the exact lines needed to fix the bug. 
However, this number increases as we increase the granularity to functions and files where we found that around half of the issues already provide the location of the file needed to be edited in the description.
To repair a bug or introduce a new feature, finding the location to make the edit is extremely important. 
As such, we leverage this classification and focus our later analysis on the effect the provided location has on the repair performance of \tech and baseline approaches.

These classification dimensions and categories raise potential issues with the \swebenchlite problems such as unsolvable questions, misleading potential solutions, and significant differences in problem difficulties.
These issues have not been properly considered by either the benchmark creation process or prior approaches. 
Furthermore, we hope our classification can provide additional insights on the type of problems that can be solved by existing and future approaches.

\subsection{\swebenchlitefiltered}

\begin{table}[h]
\footnotesize
\centering
\caption{Performance and ranking on \swebenchlitefiltered. * indicates a tie in ranking.}
\label{tab:newbenchmark}
\scalebox{0.8}{
\begin{tabular}{ll|rlrl}
\toprule
     \multirow{2}*{Tool} & \multirow{2}*{\llm} & \multicolumn{2}{c}{\swebenchlite} & \multicolumn{2}{c}{\swebenchlitefiltered} \\
     & &  \%{} Resolved & Rank & \%{} Resolved & Rank \\
\midrule

\cellcolor[HTML]{e1ecff}\codestoryaide{}~\cite{codestoryaide} \lock{} & \cellcolor[HTML]{e1ecff}\makecell[l]{\openai{} \gptfouro{}+\anthropic{} \claudesonnet} & \cellcolor[HTML]{e1ecff}129 (43.00\%) & \cellcolor[HTML]{e1ecff}1 & \cellcolor[HTML]{e1ecff}114 (45.78\%) &\cellcolor[HTML]{e1ecff}1\\
\bytedanceagent{}~\cite{marscode} \lock{} & NA & 118 (39.33\%) & 2 & 106 (42.57\%) &2\\
\cellcolor[HTML]{e1ecff}\honeycomb{}~\cite{honeycomb} \lock{} & \cellcolor[HTML]{e1ecff}NA & \cellcolor[HTML]{e1ecff}115 (38.33\%) & \cellcolor[HTML]{e1ecff}3 & \cellcolor[HTML]{e1ecff}98 (39.36\%) &\cellcolor[HTML]{e1ecff}3\\
\mentatbot{}~\cite{mentatbot} \lock{} & \openai{} \gptfouro{} & 114 (38.00\%) & 4 & 96 (38.55\%) &4\\
\cellcolor[HTML]{e1ecff}\gru{}~\cite{gru} \lock{} & \cellcolor[HTML]{e1ecff}NA & \cellcolor[HTML]{e1ecff}107 (35.67\%) & \cellcolor[HTML]{e1ecff}5 & \cellcolor[HTML]{e1ecff}94 (37.75\%) &\cellcolor[HTML]{e1ecff}5\\
\isoform{}~\cite{isoform} \lock{} & NA & 105 (35.00\%) & 6 & 91 (36.55\%) &6\\
\cellcolor[HTML]{e1ecff}\supercoder{}~\cite{supercoder} \lock{} & \cellcolor[HTML]{e1ecff}NA & \cellcolor[HTML]{e1ecff}102 (34.00\%) & \cellcolor[HTML]{e1ecff}7 & \cellcolor[HTML]{e1ecff}87 (34.94\%) &\cellcolor[HTML]{e1ecff}7\\
\lingma{}~\cite{lingma} \lock{} & \makecell[l]{\openai{} \gptfouro{}+\anthropic{} \claudesonnet} & 99 (33.00\%) & 8 & 86 (34.54\%) &8\\
\cellcolor[HTML]{e1ecff}\factorydroid{}~\cite{factorydroid} \lock{} & \cellcolor[HTML]{e1ecff}NA & \cellcolor[HTML]{e1ecff}94 (31.33\%) & \cellcolor[HTML]{e1ecff}10 & \cellcolor[HTML]{e1ecff}82 (32.93\%) &\cellcolor[HTML]{e1ecff}10\\
\amazonqagent{}-v2~\cite{amazonqdeveloper} \lock{} & NA & 89 (29.67\%) & 12* & 76 (30.52\%) &13\\
\cellcolor[HTML]{e1ecff}\coder{}~\cite{coder} \lock{} & \cellcolor[HTML]{e1ecff}\openai{} \gptfour{} & \cellcolor[HTML]{e1ecff}85 (28.33\%) & \cellcolor[HTML]{e1ecff}14 & \cellcolor[HTML]{e1ecff}71 (28.51\%) &\cellcolor[HTML]{e1ecff}14*\\
\masai{}~\cite{masai} \lock{} & NA & 84 (28.00\%) & 15 & 70 (28.11\%) &16\\
\cellcolor[HTML]{e1ecff}\sima{}~\cite{sima} \lock{} & \cellcolor[HTML]{e1ecff}\openai{} \gptfouro{} & \cellcolor[HTML]{e1ecff}83 (27.67\%) & \cellcolor[HTML]{e1ecff}16 & \cellcolor[HTML]{e1ecff}71 (28.51\%) &\cellcolor[HTML]{e1ecff}14*\\
\ibmagent{}~\cite{ibmagent} \lock{} & NA & 80 (26.67\%) & 17* & 66 (26.51\%) &18*\\
\cellcolor[HTML]{e1ecff}\opencsgagent{}~\cite{opencsgstarship} \lock{} & \cellcolor[HTML]{e1ecff}\openai{} \gptfour{} & \cellcolor[HTML]{e1ecff}71 (23.67\%) & \cellcolor[HTML]{e1ecff}22 & \cellcolor[HTML]{e1ecff}56 (22.49\%) &\cellcolor[HTML]{e1ecff}23*\\
\amazonqagent{}~\cite{amazonqdeveloper} \lock{} & NA & 61 (20.33\%) & 26 & 51 (20.48\%) &25*\\
\cellcolor[HTML]{e1ecff}\repounderstander{}~\cite{repounderstander} \lock{} & \cellcolor[HTML]{e1ecff}\openai{} \gptfour{} & \cellcolor[HTML]{e1ecff}64 (21.33\%) & \cellcolor[HTML]{e1ecff}25 & \cellcolor[HTML]{e1ecff}51 (20.48\%) &\cellcolor[HTML]{e1ecff}25*\\
\midrule
\autocoderover{}-v2~\cite{autocoderovertwo} & \openai{} \gptfouro{} & 92 (30.67\%) & 11 & 79 (31.73\%) &11\\
\cellcolor[HTML]{e1ecff}\repograph~\cite{repograph} & \cellcolor[HTML]{e1ecff}\openai{} \gptfouro{} & \cellcolor[HTML]{e1ecff}89 (29.67\%) & \cellcolor[HTML]{e1ecff}12* & \cellcolor[HTML]{e1ecff}77 (30.92\%) &\cellcolor[HTML]{e1ecff}12\\
\multirow{1}*{\moatless{}~\cite{moatless}} & \anthropic{} \claudesonnet{} & 80 (26.67\%) & 17* & 67 (26.91\%) &17\\
 & \openai{} \gptfouro{} & 74 (24.67\%) & 21 & 62 (24.90\%) &21\\
\cellcolor[HTML]{e1ecff}\opendevincodeact{}~\cite{opendevin} & \cellcolor[HTML]{e1ecff}\anthropic{} \claudesonnet & \cellcolor[HTML]{e1ecff}80 (26.67\%) & \cellcolor[HTML]{e1ecff}17* & \cellcolor[HTML]{e1ecff}65 (26.10\%) &\cellcolor[HTML]{e1ecff}20\\
\aider{}~\cite{aidar} & \makecell[l]{\openai{} \gptfouro{}+\anthropic{} \claudesonnet} & 79 (26.33\%) & 20 & 66 (26.51\%) &18*\\
\cellcolor[HTML]{e1ecff}\multirow{1}*{\sweagent{}~\cite{sweagent}} & \cellcolor[HTML]{e1ecff}\anthropic{} \claudesonnet{} & \cellcolor[HTML]{e1ecff}69 (23.00\%) & \cellcolor[HTML]{e1ecff}23 & \cellcolor[HTML]{e1ecff}58 (23.29\%) &\cellcolor[HTML]{e1ecff}22\\
\cellcolor[HTML]{e1ecff} & \cellcolor[HTML]{e1ecff}\openai{} \gptfouro{} & \cellcolor[HTML]{e1ecff}55 (18.33\%) & \cellcolor[HTML]{e1ecff}28 & \cellcolor[HTML]{e1ecff}45 (18.07\%) &\cellcolor[HTML]{e1ecff}27*\\
\cellcolor[HTML]{e1ecff} & \cellcolor[HTML]{e1ecff}\openai{} \gptfour{} & \cellcolor[HTML]{e1ecff}54 (18.00\%) & \cellcolor[HTML]{e1ecff}29 & \cellcolor[HTML]{e1ecff}42 (16.87\%) &\cellcolor[HTML]{e1ecff}29\\
\appmapnavie{}~\cite{appmapnavie} & \openai{} \gptfouro{} & 65 (21.67\%) & 24 & 56 (22.49\%) &23*\\
\cellcolor[HTML]{e1ecff}\autocoderover{}~\cite{autocoderover} & \cellcolor[HTML]{e1ecff}\openai{} \gptfour{} & \cellcolor[HTML]{e1ecff}57 (19.00\%) & \cellcolor[HTML]{e1ecff}27 & \cellcolor[HTML]{e1ecff}45 (18.07\%) &\cellcolor[HTML]{e1ecff}27*\\
\midrule
\rag{}~\cite{sweagent} & \anthropic{} \claudeopus{} & 13 (4.33\%) & 30 & 10 (4.02\%) &30\\
 & \openai{} \gptfour{} & 8 (2.67\%) & 32 & 5 (2.01\%) &32\\
 & \anthropic{} \claude{}-2 & 9 (3.00\%) & 31 & 6 (2.41\%) &31\\
 & \openai{} \chatgpt{} & 1 (0.33\%) & 33 & 0 (0.00\%) &33\\
 \midrule
\cellcolor[HTML]{e1ecff}\textbf{\tech{}} & \cellcolor[HTML]{e1ecff}\openai{} \gptfouro{} & \cellcolor[HTML]{e1ecff}96 (32.00\%) & \cellcolor[HTML]{e1ecff}9 & \cellcolor[HTML]{e1ecff}84 (33.73\%) &\cellcolor[HTML]{e1ecff}9\\

\bottomrule
\end{tabular}
}
\end{table}

Building on the above problem classifications, we will more rigorously compare and contrast \tech and existing work. 
Specifically, we focus on a subset of the problems in \swebenchlite after removing the problems that contain the exact patch in the problem description, misleading solutions, or do not provide enough information in the original issue description.  
This eliminates the less reasonable problems and normalizes the difficulty level of the benchmark. 
We refer to our subset of \sanitizedtotalproblem{} problems as \swebenchlitefiltered{}. 
We note here that our approach of identifying and excluding problematic problems has already been confirmed by a later work done by OpenAI, where they have released a similar filtered benchmark of \swebenchverified~\cite{openaiverified}.

Table~\ref{tab:newbenchmark} shows the results on the \swebenchlitefiltered{} benchmark and the corresponding ranking of each approach.
We also included the results on the original 300 problems in \swebenchlite for comparison. 
While the general ranking of all approaches stay roughly the same, we do observe some small ranking changes.
Compared to the original \swebenchlite, our filtered benchmark of \swebenchlitefiltered provides a more accurate reflection of the true capability of autonomous software development tools.

\begin{figure*}[h]
\centering
    \begin{subfigure}[b]{0.32\textwidth}
        \centering
        \includegraphics[width=\textwidth]{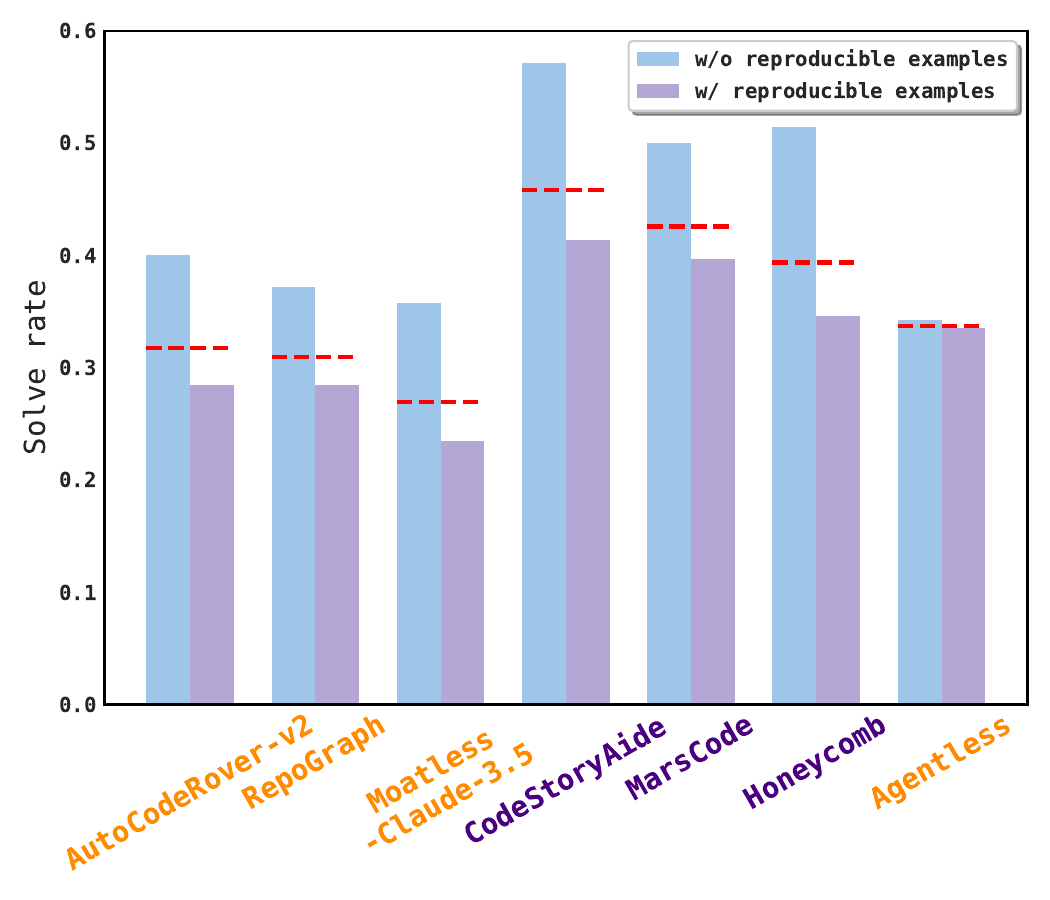}
        \caption{Description quality}
        \label{fig:filered_benchmark_description}
    \end{subfigure}
    \begin{subfigure}[b]{0.32\textwidth}
        \centering
        \includegraphics[width=\textwidth]{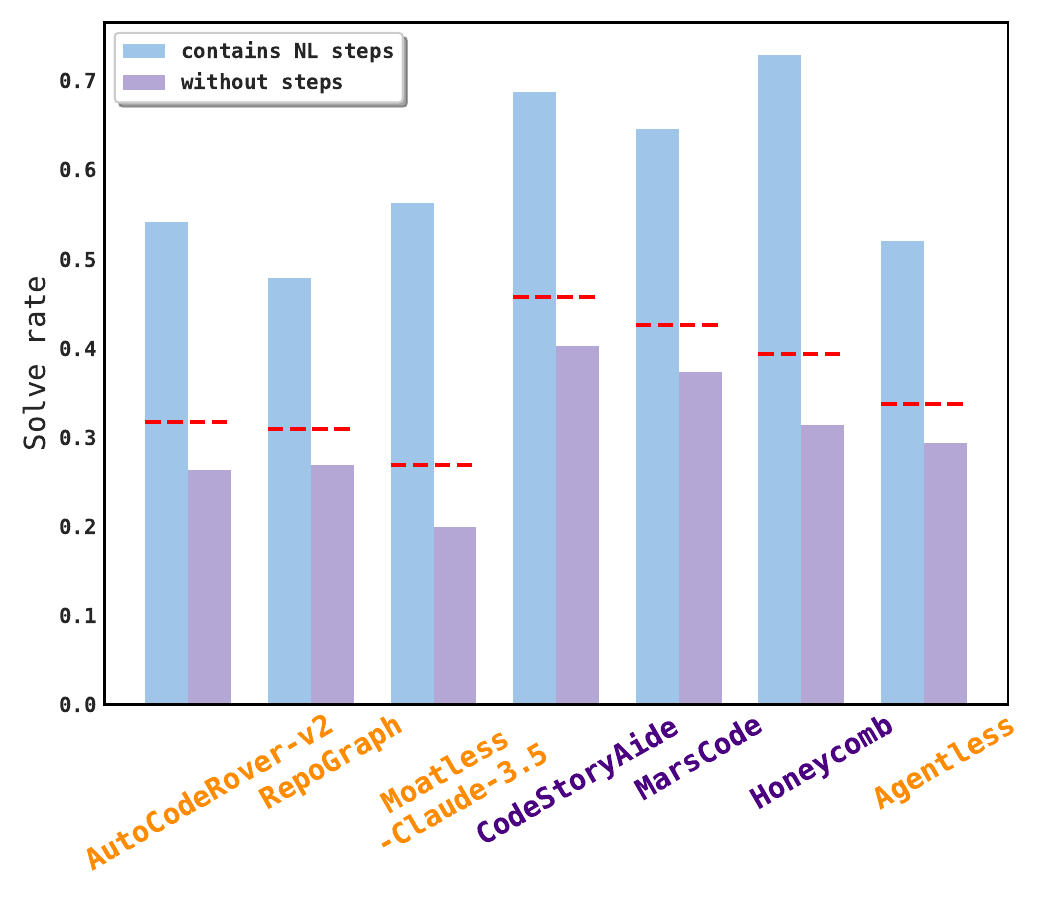}
        \caption{Solution in description}
        \label{fig:filered_benchmark_patch}
    \end{subfigure}
    \begin{subfigure}[b]{0.32\textwidth}
        \centering
        \includegraphics[width=\textwidth]{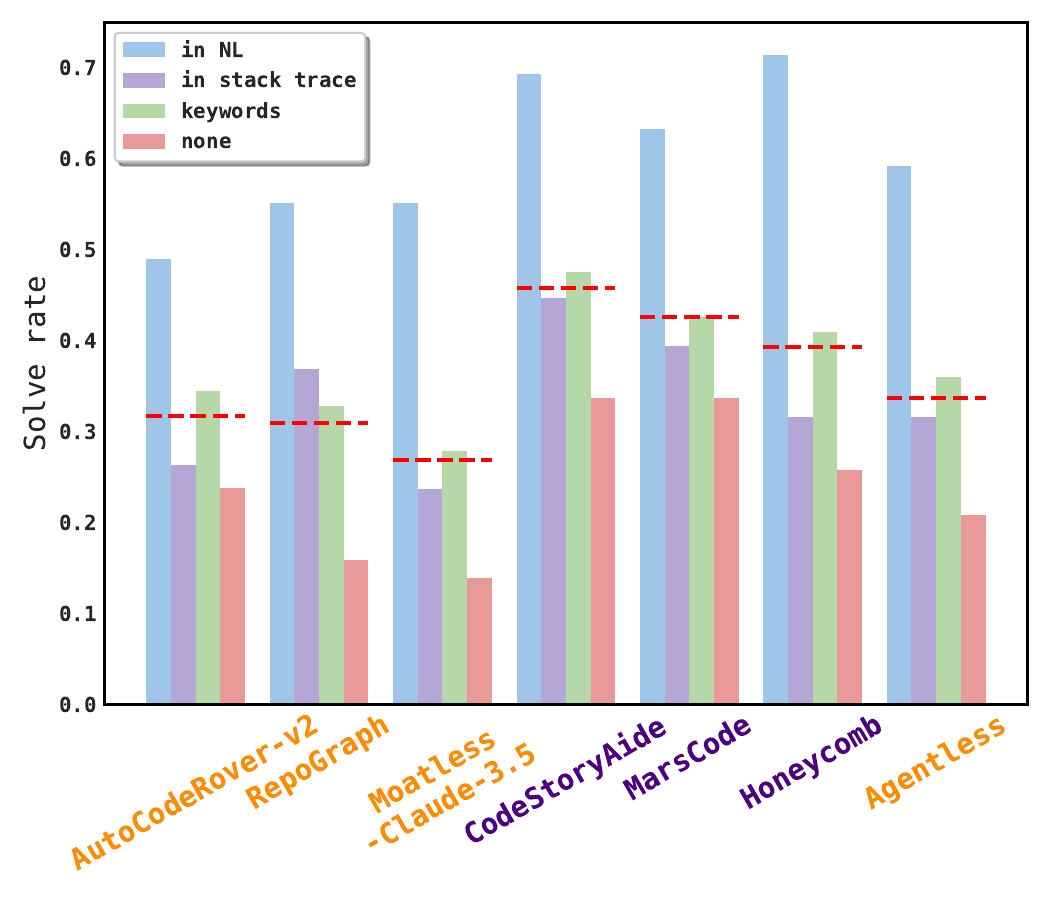}
        \caption{Location information}
        \label{fig:filtered_benchmark_location}
    \end{subfigure}
    \caption{Solve rate of selected approaches (\textbf{\textcolor[HTML]{db8500}{orange}} means open-source while \textbf{\textcolor[HTML]{2e0063}{indigo}} means closed-source) on different problem categories in \swebenchlitefiltered{}. \textcolor{red}{Red} dotted line indicates the average solve rate on the entire \swebenchlitefiltered for each approach.
    } 
    \label{fig:filtered_comparison}
\end{figure*}

Using the classification results, we further examine the types of problems that are solved by \tech and prior approaches on \swebenchlitefiltered. 
Figure~\ref{fig:filtered_comparison} shows the solve rate of various top-performing open-source and closed-source approaches across the different categories of problems. 
We first examine if having code examples to reproduce the error in the issue description can help the \llm better solve the issue in Figure~\ref{fig:filered_benchmark_description}.
Surprisingly, we found that the solve rate of all prior approaches drop when evaluated on the problems with reproducible code examples.
Many agent-based approaches~\cite{sweagent, opendevin, coder} attempt to first reproduce the error, however, this may not improve performance even on problems with already provided reproducible examples. 
However, we observe the performance for \tech remains very high on the problems with reproducible code examples. 
This is because \tech generates \reproduction tests using the original issues descriptions, hence can better make use of the reproducible code examples provided (similarly for \mentatbot{} which also contains an explicit test generation step).
This demonstrate the importance of the test generation stage for patch selection.
Next, we look at the effect of ground truth patch/solution in the issue description. 
Figure~\ref{fig:filered_benchmark_patch} shows the expected result where all selected techniques perform better on issues that already provide solution steps in natural language. 
Furthermore, in Figure~\ref{fig:filtered_benchmark_location}, we examine the solve rate with respect to the location information provided in the issues description.
Unsurprisingly, we found that the highest solve rates are on problems where the location is provided in natural language followed by stack traces. 
The most difficult problems are those that do not contain any clues about the location of the issue in the description.
We observe that compared with closed-source approaches, \tech performs comparably when the location is provided in either natural language, stack trace, or keywords. 
However, the closed-source agent tools perform better compared to \tech in the case where no location clue is provided.
This highlights an advantage of agent-based tools in solving these more complex problems where they are able to use complex code search tools. 
This represents potential future work for \tech to target and further improve these types of problems.

\subsection{\swebenchverified}

\begin{table}[h]
\footnotesize
\centering
\caption{Performance on \swebenchverified.}
\label{tab:swebenchverified}
\scalebox{0.9}{
\begin{tabular}{ll|r}
\toprule
 Tool & \llm & \%{} Resolved \\
\midrule

\cellcolor[HTML]{e1dcef}\multirow{1}*{\toolstool{}~\cite{tools} \lock{}} & \cellcolor[HTML]{e1dcef}\anthropic{} \claudesonnet{}onnet (20241022) & \cellcolor[HTML]{e1dcef}245 (49.00\%) \\
\cellcolor[HTML]{e1dcef} & \cellcolor[HTML]{e1dcef}\anthropic{} \claudehaiku{} & \cellcolor[HTML]{e1dcef}203 (40.60\%) \\
\solversolver{}~\cite{solver} \lock{} & NA & 227 (45.40\%) \\
\cellcolor[HTML]{e1dcef}\gru{}~\cite{gru} \lock{} & \cellcolor[HTML]{e1dcef}NA & \cellcolor[HTML]{e1dcef}226 (45.20\%) \\
\honeycomb{}~\cite{honeycomb} \lock{} & NA & 203 (40.60\%) \\
\cellcolor[HTML]{e1dcef}\amazonqagent{}-v2~\cite{amazonqdeveloper} \lock{} & \cellcolor[HTML]{e1dcef}NA & \cellcolor[HTML]{e1dcef}194 (38.80\%) \\
\factorydroid{}~\cite{factorydroid} \lock{} & NA & 185 (37.00\%) \\
\cellcolor[HTML]{e1dcef}\amazonqagent{}~\cite{amazonqdeveloper} \lock{} & \cellcolor[HTML]{e1dcef}NA & \cellcolor[HTML]{e1dcef}128 (25.60\%) \\
\midrule
\composio{}~\cite{composio} & \anthropic{} \claudesonnet{}onnet & 203 (40.60\%) \\
\cellcolor[HTML]{e1dcef}\autocoderover{}-v2~\cite{autocoderovertwo} & \cellcolor[HTML]{e1dcef}\openai{} \gptfouro{} & \cellcolor[HTML]{e1dcef}192 (38.40\%) \\
\multirow{1}*{\sweagent{}~\cite{sweagent}} & \anthropic{} \claudesonnet{}onnet & 168 (33.60\%) \\
 & \openai{} \gptfouro{} & 116 (23.20\%) \\
 & \openai{} \gptfour{} & 112 (22.40\%) \\
\cellcolor[HTML]{e1dcef}\lingma{}~\cite{lingma} & \cellcolor[HTML]{e1dcef}\lingmaswegpt & \cellcolor[HTML]{e1dcef}144 (28.80\%) \\
\appmapnavie{}~\cite{appmapnavie} & \openai{} \gptfouro{} & 131 (26.20\%) \\
\midrule
\textbf{\tech} & \openai{} \gptfouro{} & 194 (38.80\%) \\ 
\bottomrule
\end{tabular}
}
\end{table}

Similar to \swebenchlitefiltered{} and inspired by similar concerns in Section~\ref{sec:benchmark_analysis}, OpenAI has produced a newly filtered dataset of \swebenchverified~\cite{openaiverified} validated by human developers to ensure each issue has sufficient amounts of information to be solved. 
Table~\ref{tab:swebenchverified} shows the performance of \tech compared with prior agent-based approaches on \swebenchverified. 
Similar to the results on \swebenchlite, \tech maintains its strong performance and is able to solve 194 out of 500 problems (38.80\%). 
\tech is able to achieve the second highest performance compared with open-source approaches and perform better than many closed-source / commercial techniques. 
Furthermore, we note that \tech performs the best among all techniques that use \gptfouro as the \llm. 

%% file: secs/conclusion.tex
\section{Threats to Validity}

\parabf{Internal.} 
One threat to validity comes from the data leakage of ground truth developer patches in \swebenchlite being part of the training data for \gptfouro. 
Since \gptfouro is a closed-source model, we do not have access to the training data.
Meanwhile, we note here that prior work almost exclusively used similar closed-source \llm{s} (e.g., \gptfouro, \gptfour, \claude-3.5, etc), and our approach can outperform all existing open-source solutions with same models.
Furthermore, the authors of \swebench~\cite{swebench} compared the resolve rate of issues collected before and after the knowledge cutoff date of \gptfour, and did not find any significant difference.  
To completely address this threat, we would need to retrain \gptfouro from scratch which would be infeasible for an academic project.

\parabf{External.}
One main external threat comes from our evaluation dataset of \swebenchlite.
While the performance of \tech might not generalize to other datasets, \swebenchlite is by far the most popular evaluation dataset which contains a diverse range of problems.
In addition, \emph{OpenAI has independently performed an extensive evaluation of \tech} and other open-source solutions on \swebenchlite, \swebench, and their new \swebenchverified benchmark, further confirming that \tech outperforms all other open-source agents~\cite{openaiverified}. Moreover, on Sept. 12th 2024, OpenAI just released the new o1 model family and also adopted Agentless as the top approach to showcase their performance on \swebench~\cite{openaioonesystemcard}.
In the future, we plan to further address this threat by evaluating \tech on other benchmarks.

\section{Conclusion}

We propose \tech -- an \emph{agentless} approach to automatically tackle software development problems. 
\tech uses a simple three phase approach of localization, repair, and patch validation. 
Compared to prior agent-based approaches, \tech deliberately disallows the \llm{ for autonomous tool usage or planning}. 
Our evaluation on the popular \swebenchlite benchmark demonstrates that \tech can achieve the highest performance compared with other open-source techniques while at the same time minimizing the cost. 
Furthermore, we perform a detailed classification of problems in \swebenchlite to not only offer new insights but to construct a more rigorous benchmark of \swebenchlitefiltered{} after removing problematic problems.

\section*{Acknowledgments}

We thank Jiawei Liu and Yuxiang Wei for providing some of the{ resources} used to run the experiments. One of the authors would like to thank Jun Yang for generously gifting his old bike\footnote{Sadly the bike is currently broken.} which allowed the author to travel faster and thus increasing research speed.